\documentclass[aps,prd,twocolumn,groupedaddress,nofootinbib]{revtex4}
\usepackage{graphicx,color}

\usepackage[english]{babel}   
\usepackage{amssymb}
\usepackage{amsmath}
\usepackage{amsfonts}
\usepackage{upgreek}
\usepackage{latexsym}
\usepackage{booktabs}
\usepackage{dcolumn} 

\newcommand{\Stu}{St\"uckelberg\,\,}

\newcommand{\Fmn}{F_{\mu\nu}}
\newcommand{\FMN}{F^{\mu\nu}}
\newcommand{\Am}{A_{\mu}}

\renewcommand\({\left(}
\renewcommand\){\right)}
\renewcommand\[{\left[}
\renewcommand\]{\right]}

\newcommand\n{{\mbox {\boldmath $\nabla$}}}
\newcommand{\ra}{\rightarrow}

\def\lsim{\raise 0.4ex\hbox{$<$}\kern -0.8em\lower 0.62
ex\hbox{$\sim$}}

\def\gsim{\raise 0.4ex\hbox{$>$}\kern -0.7em\lower 0.62
ex\hbox{$\sim$}}

\def\lbar{{\hbox{$\lambda$}\kern -0.7em\raise 0.6ex
\hbox{$-$}}}

\newcommand\eq[1]{eq.~(\ref{#1})}
\newcommand\eqs[2]{eqs.~(\ref{#1}) and (\ref{#2})}
\newcommand\Eq[1]{Equation~(\ref{#1})}
\newcommand\Eqs[2]{Equations~(\ref{#1}) and (\ref{#2})}

\newcommand\eqst[2]{eqs.~(\ref{#1})--(\ref{#2})}
\newcommand\Eqst[2]{Eqs.~(\ref{#1})--(\ref{#2})}
\newcommand\pa{\partial}
\newcommand\p{\partial}

\newcommand\ee{\end{equation}}
\newcommand\be{\begin{equation}}
\def\bea{\begin{array}}
\def\eea{\end{array}}\def\ea{\end{array}}
\newcommand\ees{\end{eqnarray}}
\newcommand\bees{\begin{eqnarray}}
\def\nn{\nonumber}

\def\a{\alpha}
\def\b{\beta}

\def\g{\gamma}

\def\d{\delta}

\def\eps{\epsilon}

\def\dslash{\hspace{-1mm}\not{\hbox{\kern-2pt $\partial$}}}
\def\Dslash{\not{\hbox{\kern-4pt $D$}}}
\def\pslash{\not{\hbox{\kern-2.1pt $p$}}}
\def\kslash{\not{\hbox{\kern-2.3pt $k$}}}
\def\qslash{\not{\hbox{\kern-2.3pt $q$}}}

\newcommand{\vk}{{\bf k}}
\newcommand{\vx}{{\bf x}}

\def\p1{{\bf p}_1}
\def\p2{{\bf p}_2}
\def\k1{{\bf k}_1}
\def\k2{{\bf k}_2}

\newcommand{\emn}{\eta_{\mu\nu}}

\newcommand{\eMN}{\eta^{\mu\nu}}
\newcommand{\eRS}{\eta^{\rho\sigma}}
\newcommand{\eMR}{\eta^{\mu\rho}}
\newcommand{\eNS}{\eta^{\nu\sigma}}
\newcommand{\eMS}{\eta^{\mu\sigma}}
\newcommand{\eNR}{\eta^{\nu\rho}}

\newcommand{\gmn}{g_{\mu\nu}}

\newcommand{\gMN}{g^{\mu\nu}}

\newcommand{\gBmn}{\bar{g}_{\mu\nu}}

\newcommand{\hmn}{h_{\mu\nu}}
\newcommand{\hrs}{h_{\rho\sigma}}
\newcommand{\hmr}{h_{\mu\rho}}

\newcommand{\hrn}{h_{\rho\nu}}

\newcommand{\hMN}{h^{\mu\nu}}
\newcommand{\hRS}{h^{\rho\sigma}}

\newcommand{\hij}{h_{ij}}

\newcommand{\hTTij}{h_{ij}^{\rm TT}}

\newcommand{\xm}{x^{\mu}}
\newcommand{\xn}{x^{\nu}}

\newcommand{\xim}{\xi_{\mu}}
\newcommand{\xin}{\xi_{\nu}}

\newcommand{\pam}{\pa_{\mu}}

\newcommand{\pan}{\pa_{\nu}}
\newcommand{\parho}{\pa_{\rho}}

\newcommand{\paM}{\pa^{\mu}}
\newcommand{\paN}{\pa^{\nu}}
\newcommand{\paR}{\pa^{\rho}}
\newcommand{\paS}{\pa^{\sigma}}

\newcommand{\Tmn}{T_{\mu\nu}}

\newcommand{\TMN}{T^{\mu\nu}}

\newcommand{\dddM}{\kern 0.2em \raise 1.9ex\hbox{$...$}\kern -1.0em \hbox{$M$}}
\newcommand{\dddQ}{\kern 0.2em \raise 1.9ex\hbox{$...$}\kern -1.0em \hbox{$Q$}}
\newcommand{\dddI}{\kern 0.2em \raise 1.9ex\hbox{$...$}\kern -1.0em\hbox{$I$}}
\newcommand{\dddJ}{\kern 0.2em \raise 1.9ex\hbox{$...$}\kern-1.0em
\hbox{$J$}}
\newcommand{\dddcalJ}{\kern 0.2em \raise 1.9ex\hbox{$...$}\kern-1.0em
\hbox{${\cal J}$}}

\newcommand{\dddO}{\kern 0.2em \raise 1.9ex\hbox{$...$}\kern -1.0em
\hbox{${\cal O}$}}
\def\dddz{\raise 1.5ex\hbox{$...$}\kern -0.8em \hbox{$z$}}
\def\dddd{\raise 1.8ex\hbox{$...$}\kern -0.8em \hbox{$d$}}
\def\dddbd{\raise 1.8ex\hbox{$...$}\kern -0.8em \hbox{${\bf d}$}}
\def\ddbd{\raise 1.8ex\hbox{$..$}\kern -0.8em \hbox{${\bf d}$}}
\def\dddx{\raise 1.6ex\hbox{$...$}\kern -0.8em \hbox{$x$}}

\begin{document}

\title{ Bardeen variables and hidden gauge symmetries\\  in linearized massive gravity}

\author{Maud Jaccard}
\author{Michele Maggiore}
\author{Ermis Mitsou}
\affiliation{D\'epartement de Physique Th\'eorique and Center for Astroparticle Physics, 
Universit\'e de Gen\`eve, 24 quai Ansermet, CH--1211 Gen\`eve 4, Switzerland}

\email{maud.jaccard@unige.ch, michele.maggiore@unige.ch, ermis.mitsou@unige.ch}
%
%
%
%

\begin{abstract}

We give a detailed discussion of the use  of the $(3+1)$ decomposition  and of Bardeen's variables in   massive gravity linearized over a Minkowski as well as over a de~Sitter background.  In Minkowski space the Bardeen ``potential" $\Phi$, that in the massless case is a non-radiative degree of freedom, becomes radiative and describes the helicity-$0$ component of the massive graviton. Its dynamics is governed by a simple Klein-Gordon action, supplemented by a term $(\Box\Phi)^2$ if we do not make the Fierz-Pauli tuning of the mass term. In de~Sitter the identification of the variable that describes the radiative degree of freedom in the scalar sector is more subtle, and even involves  expressions non-local in time. The use of this new variable provides a simple and transparent derivation of the Higuchi bound and of the disappearance of the scalar degree of freedom at a special value of $m_g^2/H^2$.

The use of this formalism  also allows us to  uncover the existence of a hidden gauge symmetry of the massive theory, that becomes manifest  only once the non-dynamical components of the metric are integrated out, and that  is present both in Minkowski and in de~Sitter.

\end{abstract}


\maketitle

\section{Introduction}

The problem of formulating  a consistent theory of gravity with a massive  graviton has a long history. Already in 1939  Fierz and Pauli \cite{Fierz:1939ix} showed that, when the theory is linearized over Minkowski space, a specific form for the mass term is required to avoid the appearance of a sixth ghost-like degree of freedom. In the 1970s, in another classic paper, Boulware and Deser \cite{Boulware:1973my} showed that even with the Fierz-Pauli (FP) mass term the ghost reappears when the non-linearities of the gravitational field are taken into account. Recent years have witnessed a flurry of activity on the problem, stimulated  by the possible relevance of infrared modifications of gravity for understanding the origin of dark energy, and 
by theoretical breakthroughs \cite{ArkaniHamed:2002sp,deRham:2010ik,deRham:2010kj}
that also have an  intrinsic field-theoretical interest. 

A recurrent theme in the study of both gauge theories and general relativity   is that a good choice of variables can significantly enlighten the physics, and different choices are appropriate for studying different aspects of the theory. Indeed,  the recent breakthroughs in massive gravity can be partly traced to a clever way of isolating the dynamics of the helicity-$0$ mode through a generalization of the {\Stu} trick to general relativity, proposed in 
\cite{ArkaniHamed:2002sp} (and further studied in various contexts in 
\cite{Luty:2003vm,Nicolis:2004qq,Dubovsky:2004sg,Creminelli:2005qk,Deffayet:2005ys}.
See \cite{Hinterbichler:2011tt} for a recent review).
This lead recently to the construction of a consistent ghost-free theory of massive gravity  to all orders in the decoupling limit and up to quartic order in the non-linearities away from the decoupling limit
\cite{deRham:2010ik,deRham:2010kj,deRham:2011rn}. The absence of  ghosts in this model was proved in full generality using the ADM formalism in \cite{Hassan:2011hr}.
Ghost-free actions with a general reference metric that admits massive spin-2 fluctuations around non-flat backgrounds were first considered in \cite{Hassan:2011vm}, and were proven to be free of the Boulware-Deser ghost in \cite{Hassan:2011tf,Hassan:2011ea,Hassan:2012qv}.
Ghost-free dynamics for the reference metric is presented  in \cite{Hassan:2011zd}.
Further recent work on ghost-free bimetric theories includes \cite{Hassan:2012wr,Hassan:2012gz,Comelli:2012db,Berg:2012kn}.

In this paper we provide a detailed discussion of the use of the (3+1)  decomposition of the metric and of the gauge-invariant Bardeen variables in linearized massive gravity.  This formalism is  a standard tool of cosmological perturbation theory \cite{Mukhanov:1990me}. In massive gravity linearized over Minkowski space it was first introduced in  
\cite{Deser:1966zzb}, and 
has been applied to massive gravity in a number of recent papers, see e.g.~\cite{Rubakov:2008nh,Alberte:2010it,Alberte:2011ah}.
Elaborating on these results, 
we will see that these variables can be quite useful for elucidating various aspects of the massive theory.   In particular we will see that in Minkowski space, as the mass term approaches the FP form, after elimination of the non-dynamical components of the metric the dynamics in the scalar sector nicely collapses to a massive Klein-Gordon action for the Bardeen variable $\Phi$, which therefore describes the helicity-$0$ component of the massive graviton (see also \cite{Deser:1966zzb,Alberte:2010it}). It is quite interesting to see that $\Phi$, which in the massless case describes a non-radiative degree of freedom (and is in fact usually called the Bardeen ``potential") becomes a radiative degree of freedom when the FP mass term is switched on. Furthermore, for a generic mass term the action in the scalar sector can be reduced to a higher-derivative theory for $\Phi$, in which only Lorentz-covariant structures such as $(\Box\Phi)^2$ appear. More generally, we will see that the use of the Bardeen variable $\Phi$ can be a convenient way of isolating the dynamics of the helicity-0  mode, complementary to the by now standard {\Stu} formalism.

An intriguing consequence of the fact that the helicity-0 mode of the massive graviton is described by  $\Phi$ is that a residual gauge symmetry  appears, consisting of the transformations 
$\hmn\ra\hmn  -(\pam\xin+\pan\xim)$ with $\xim$ parametrized by two scalar functions $A(x)$ and $C(x)$ as
$\xi_0=A$ and $\xi_i=\pa_i C$. We will see that this symmetry  only appears when one eliminates  the non-dynamical components of the metric,
remaining just with the  five independent fields that describe the  physical components of a massive graviton
(plus the extra ghost-like scalar if we are  away from the FP point).

We will then turn to massive gravity linearized over a de~Sitter background. We will see that in this case the identification of the radiative degree of freedom that describes the helicity-0 mode of the massive graviton is quite subtle, and we will show that it even involves integrals over time of some metric components. In terms of this variable, after elimination of the non-dynamical degrees of freedom, the dynamics in the scalar sector of linearized massive gravity in de~Sitter (with FP mass term) collapses again to a simple Klein-Gordon action. The overall coefficient of this action nicely displays the existence of the Higuchi bound as well as of a point of enhanced symmetry corresponding to the so-called ``partially massless" theory, where the scalar degree of freedom disappears.
At the same time this dynamical variable is gauge-invariant, which shows that even in de~Sitter  space, after elimination of the non-dynamical fields, gauge transformations with $\xi_0=A$ and $\xi_i=\pa_i C$ are a symmetry of the massive theory.

The paper is organized as follows. In sect.~\ref{sect:harmo} we recall the (3+1) decomposition of the metric in flat space-time and  we illustrate how it can be used to separate the metric into  pure gauge, radiative and non-radiative degrees of freedom. We also discuss the behavior  under Lorentz transformations of the  variables entering the $(3+1)$ decomposition and of the gauge-invariant Bardeen's variables. Most of the  material in this section is known in the literature (except for the part on Lorentz transformations of the variables entering the (3+1) decomposition), but we find useful to present in a systematic way various  results that will be needed in the rest of the paper.
In sect.~\ref{sect:generic} we use these variables to study massive gravity, linearized over Minkowski space, for a generic quadratic mass term. We explicitly identify  the ghost degree of freedom and we find  that, at the FP point, the scalar sector is described by the Bardeen variable $\Phi$. We will then see that, outside the FP point, the scalar sector can be reduced to a simple higher-derivative theory for $\Phi$. The hidden gauge symmetry that emerges from this analysis is discussed in sect.~\ref{sect:hidden}. In sect.~\ref{sect:Stu} we discuss this symmetry from the point of view of the {\Stu} formalism.
In sect.~\ref{sect:electro} we compare our results with a similar analysis performed in massless and massive electrodynamics.  In sect.~\ref{sect:DS}
we  discuss massive gravity
linearized  over a  de~Sitter background. 
Sect.~\ref{sect:concl} contains our conclusions and a summary of the main results. Some technical material is relegated in  appendixes. 
We use the signature $\emn =(-,+,+,+)$, units $\hbar=c=1$, and we define
$\kappa \equiv (32\pi G)^{1/2}$. 

\section{Decomposition of the metric into pure gauge, radiative and non-radiative degrees of freedom}\label{sect:harmo}

\subsection{(3+1) decomposition of the metric}\label{sect:hardeco}

We begin with a review of the (3+1) decomposition in the massless case.
Expanding the metric $\gmn$ over flat space,  $\gmn=\emn+\hmn$,  the metric perturbation can be decomposed as
\bees
h_{00}&=&2\psi\, ,\label{hijpsi}\\
h_{0i}&=&\beta_i+\pa_i\gamma\label{h0ibetagamma}\\
h_{ij}&=&-2\phi\d_{ij}+\(\pa_i\pa_j-\frac{1}{3}\d_{ij}\n^2\)\lambda \nn \\
 & & +\frac{1}{2}(\pa_i\eps_j+\pa_j\eps_i)
 +h_{ij}^{\rm TT}\, ,\label{hijphilambdaeps}
\ees
(where $\n^2$ is the flat-space Laplacian) with the constraints
\bees
\hspace*{-5mm}\pa_i\beta^i=0\, ,\quad
\pa_i\eps^i=0\, ,\quad 
\pa^jh_{ij}^{\rm TT}=0\, ,\quad
\d^{ij}h_{ij}^{\rm TT}=0\, ,
\ees
and  the boundary conditions 
\be
\gamma\ra 0\, ,\quad \lambda\ra 0\, ,\quad \n^2\lambda\ra 0\,,\quad \eps_i\ra 0\, ,
\ee
at spatial infinity. As we will review below, these boundary conditions ensure the uniqueness of the 
decomposition~\cite{Flanagan:2005yc}. The tensor  $h_{ij}^{\rm TT}$ is a transverse traceless (TT)  $3\times 3$ symmetric tensor, so it carries two degrees of freedom. The variables
$\beta^i$ and $\eps^i$ are transverse vector fields, so they carry two degrees of freedom each, and  describe vector perturbations of the  background.  Finally, we have four  fields $\psi, \phi,\gamma, \lambda$ that are scalars under spatial rotations and describe the scalar perturbations of the background, for a total of 10 degrees of freedom.
This parametrization decomposes the metric perturbations into irreducible representations of translations and spatial rotations and corresponds to  a decomposition into eigenfunctions of the Laplacian, i.e. to harmonic analysis.  For this reason it is also called the
harmonic decomposition, and we will also refer to the variables
$\{\psi, \phi,\gamma, \lambda,\beta^i,\eps^i,h_{ij}^{\rm TT}\}$ as harmonic variables.

The linearized massless theory is invariant 
under the gauge transformations 
\be\label{1hmn}
\hmn\ra\hmn  -(\pam\xin+\pan\xim)\, ,
\ee
which corresponds to linearized diffeomorphisms.
To understand the properties of the harmonic variables  under this gauge transformation  it is useful to write the gauge functions $\xi_{\mu}$ in the form
\be\label{gaugetran}
\xi_0=A\, ,\qquad  \xi_i=B_i+\pa_iC\, , 
\ee
where $B_i$ is a transverse vector, $\pa_i B^i=0$. In terms of these variables \eq{1hmn}
reads~\cite{Flanagan:2005yc}
\bees
&\psi\ra \psi -\dot{A}\, ,\qquad
&\phi\ra \phi+\frac{1}{3}\n^2C\, ,\label{tr1}\\
&\gamma\ra\gamma -A-\dot{C}\, ,\qquad
&\lambda\ra\lambda -2C\, ,\label{tr2}\\
&\beta_i\ra\beta_i-\dot{B}_i\, ,\qquad
&\eps_i\ra\eps_i-2B_i\, ,\label{tr3}
\ees
while $h_{ij}^{\rm TT}$ is gauge invariant. As dictated by symmetry, the transformation of the scalars $\psi, \phi, \gamma$ and $\lambda$ depends only on the scalar functions
$A$ and $C$, while the transformation of the 
transverse vector fields $\beta_i$ and $\eps_i$  only depends on the transverse vector field 
$B_i$. The fact that $h_{ij}^{\rm TT}$ is gauge invariant is a consequence of the fact that, from the point of view of spatial rotations, $\xi^{\mu}$ decomposes into a 
spin-0 and a spin-1 part, while a traceless symmetric tensor such as $h_{ij}^{\rm TT}$ is a spin-2 operator.

Using the above variables  one can form the  following  gauge-invariant scalar combinations
\bees
\Phi&=&-\phi-\frac{1}{6}\n^2\lambda\, ,\label{Bard1}\\
\Psi &=& \psi -\dot{\gamma}+\frac{1}{2}\ddot{\lambda}\, ,\label{Bard2}
\ees
whose generalization to perturbations of FRW space-time gives the standard  Bardeen variables. We will still use this nomenclature in the flat-space case.
Similarly we can form a 
gauge-invariant transverse vector 
\be
\Xi_i=\beta_i-\frac{1}{2}\dot{\eps}_i\, .
\ee
Thus, at the level of linearized theory, we have six gauge-invariant quantities: the two components of the transverse traceless tensor perturbations $h_{ij}^{\rm TT}$, the two components of the vector perturbation $\Xi_i$ subject to the condition $\pa_i\Xi^i=0$, and the two scalar perturbations $\Phi$ and $\Psi$. So, the four gauge functions $\xi^{\mu}$ allow us to eliminate four pure-gauge degrees of freedom from the ten components of $\hmn$, remaining with six gauge-invariant degrees of freedom. 

It is important to appreciate that, in a generic gauge, the harmonic variables are in general   non-local functions of the metric~\cite{Flanagan:2005yc}. This can be seen inverting \eqst{hijpsi}{hijphilambdaeps}, as follows.
The variable $\psi$ is simply given by \eq{hijpsi}, while $\phi$ is obtained taking the contraction of 
\eq{hijphilambdaeps} with $\d^{ij}$,
\be\label{Lorpsiphi}
\psi =\frac{1}{2}h_{00}\, ,\qquad \phi = -\frac{1}{6}h^i_i\, .
\ee
Thus, these quantities are local functions of the metric. All the other variables, in contrast, have a non-local  dependence on  $h_{0i}$ or on $h_{ij}$. To extract  $\gamma$ we take the  divergence of 
\eq{h0ibetagamma} and we invert the Laplacian (which, with the boundary condition that $\gamma$ vanishes at infinity, is a well-defined operation). This gives
\be\label{invgamma}
\gamma =\n^{-2}(\pa^ih_{0i})\, .
\ee
To extract $\n^2\lambda$ we apply the operator $\pa^i\pa^j$ to \eq{hijphilambdaeps} and we get
\be\label{n2la}
\n^2\lambda = -\frac{1}{2}h^i_i+\frac{3}{2}\n^{-2}(\pa^i\pa^jh_{ij})\, ,
\ee
where we used the boundary condition that $\n^2\lambda$ vanishes at infinity to invert $\n^2(\n^2\lambda)$. Requiring further that $\lambda$ itself vanishes at infinity allows one to invert once more the Laplacian in \eq{n2la}
and obtain $\lambda$.  From these expressions for  $\phi$ and $\n^2\lambda$ we find that the Bardeen variable $\Phi$ can be written as
\be\label{4PhiNL}
4\Phi =h^i_i-\n^{-2}(\pa^i\pa^jh_{ij})\, .
\ee
Thus, even $\Phi$ is a non-local function of  $\hmn$, and the same holds for $\Psi$, which involves $\lambda$ and therefore a double inversion of the Laplacian.
In the vector sector the inversion of the harmonic decomposition gives
\bees
\beta_i&=&h_{0i}-\n^{-2}(\pa_i\pa^jh_{0j})\, ,\label{invbeta}\\
\eps_i&=&2\n^{-2}\[\pa^jh_{ij}-\pa_i\n^{-2}(\pa^k\pa^lh_{kl})\]\, ,\label{inveps}
\ees
while $\hTTij$ can be obtained from \eq{hijphilambdaeps}, using the above expressions for $\phi$, $\lambda$ and $\eps_i$, and involves both the operator $\n^{-2}$, and the operator $\n^{-2}\n^{-2}$. 
Observe  however that in the  massless theory, when $\Tmn=0$, we can use the gauge invariance to fix the TT gauge, where    $h_i^i=0$, $h_{0\mu}=0$ and  $\pa^i\hij=0$. In this gauge  the scalar and vector variables vanish, as well as the non-local terms in $\hTTij $, and
$\hij=\hTTij $.

\subsection{Radiative and non-radiative degrees of freedom}\label{sect:radiative}

\subsubsection{Action and equations of motion for the gauge-invariant variables}

To study the dynamics of the harmonic degrees of freedom we consider the linearization of the Einstein action,  expanding
$\gmn =\emn+\hmn$.
The quadratic part of the Einstein-Hilbert action and the interaction term with an external conserved energy-momentum tensor are given by 
\be\label{S2}
S_2+S_{\rm int}= \int d^4x\,\[\frac{1}{2\kappa^2} \hmn {\cal E}^{\mu\nu,\rho\sigma}\hrs
+ \frac{1}{2}  \hmn\TMN\]\, ,
\ee
where  $\kappa \equiv (32\pi G)^{1/2}$ and
${\cal E}^{\mu\nu,\rho\sigma}$ is the Lichnerowicz operator, defined as\footnote{Beware that, in the literature, different   conventions are used  for the overall sign of the Lichnerowicz operator. With our convention ${\cal E}^{\mu\nu,\rho\sigma}\hrs=+\Box\hMN-\ldots$.}
\bees
{\cal E}^{\mu\nu,\rho\sigma} &\equiv&
\frac{1}{2}(\eMR\eNS+\eMS\eNR-2\eMN\eRS) \Box \nn \\
 &-&\frac{1}{2}\( \eMR\paS\paN+\eNR\paS\paM+\eMS\paR\paN+\eNS\paR\paM\)  \nn \\&+& (\eRS\paM\paN+\eMN\paR\paS) 
 \, , 
\label{defE}
\ees
and $\Box=\eMN\pam\pan$ is the flat-space d'Alembertian. 
It is useful to perform the harmonic decomposition also in  the energy-momentum tensor, writing \cite{Flanagan:2005yc}
\bees
T_{00}&=&\rho\, ,\label{T00}\\
T_{0i}&=&S_i+\pa_iS\, ,\label{T0i}\\
T_{ij}&=&P\d_{ij}+\(\pa_i\pa_j-\frac{1}{3}\d_{ij}\n^2\)\sigma  \nn \\
& & +\frac{1}{2}(\pa_i\sigma_j+\pa_j\sigma_i)+\sigma_{ij} \, ,
\ees
where 
\be\label{pais1}
\pa_i\sigma^i=0\,, \quad \pa_i S^i=0\, ,\quad
\pa^i\sigma_{ij}=0\, ,\quad \d^{ij}\sigma_{ij}=0\, .
\ee
The isotropic part of $T_{ij}$ is $P\d_{ij}$, where $P$ is the pressure. The remaining terms in $T_{ij}$ define the anisotropic stress tensor and depend on a scalar $\sigma$, a transverse vector $\sigma_i$ and a TT tensor $\sigma_{ij}$. As with the harmonic decomposition of the metric, the uniqueness of the decomposition is assured if we assume that $S$,
$\sigma$, $\n^2\sigma$ and $\sigma^i$ vanish at spatial infinity.

The quantities that appear in the parametrization of $\Tmn$ are not all independent, since they are related by energy-momentum conservation $\pam\TMN=0$. Imposing 
$\pam T^{\mu 0}=0$ gives
\be\label{comdT1}
\n^2S=\dot{\rho}\, ,
\ee
while $\pam T^{\mu i}=0$ gives
\be\label{pamTmui}
\frac{1}{2}\n^2\sigma^i-\dot{S}^i+\pa^i\(-\dot{S}+P+\frac{2}{3}\n^2\sigma\)
=0\, .
\ee
Since $\sigma^i$ and $S^i$ are transverse, see \eq{pais1}, applying $\pa_i$ to this equation we get the condition 
\be
\n^2[-\dot{S}+P+(2/3)\n^2\sigma]=0\, . 
\ee
We impose  the boundary conditions that 
the energy-momentum tensor vanishes at infinity (or just  that $\dot{S}$, $P$ and
$\n^2\sigma$ vanish at infinity). Using the fact that a Poisson equation $\n^2f=0$ with the boundary condition $f=0$ at infinity  only  has the solution $f(\vx)=0$, we get
\be\label{comdT2}
-\dot{S}+P+\frac{2}{3}\n^2\sigma =0\, .
\ee
Inserting this into \eq {pamTmui} we then get
\be\label{comdT3}
\n^2\sigma^i=2\dot{S}^i\, .
\ee
Observe that the vector equation (\ref{pamTmui}) separate into an equation for the transverse vector part and
one for the scalars that parametrizes the longitudinal vector part.
Thus, overall energy-momentum conservation gives two scalar conditions, (\ref{comdT1}) and
(\ref{comdT2}), and one condition between transverse vectors, \eq{comdT3}. 

We can  now write the linearized action using the $(3+1)$ decomposition of the metric perturbations,  \eqst{hijpsi}{hijphilambdaeps}.  The  result is
\bees
\hspace*{-2mm} S_2+S_{\rm int}&=&
\frac{1}{\kappa^2}\int d^4x \[ -12\dot{\Phi}^2+4\pa_i\Phi\pa^i\Phi
-8\pa_i\Phi\pa^i\Psi  \right.
\nn\\ 
&&\hspace*{6mm}\left. +  \pa_i\Xi_j\pa^i\Xi^j 
-\frac{1}{2}\pam h_{ij}^{\rm TT}\paM h^{ij,\rm TT}\]\nn\\
&&\hspace*{-6mm} + \int d^4x \[
3\Phi P+\Psi\rho-\Xi_i S^i+\frac{1}{2}h_{ij}^{\rm TT}\sigma^{ij}\].
\label{S2Sintmzro}
\ees
As expected, the action depends only on the gauge-invariant combinations
$\Phi, \Psi,\Xi_i$ and $h_{ij}^{\rm TT}$.
We can now derive the equations of motion. Using $\{\phi,\psi,\lambda,\gamma\}$ as independent variables, 
the variation with respect to $\phi$ gives
\be\label{24ddotPhi}
24\ddot{\Phi}-8\n^2\Phi +8\n^2\Psi =-3\kappa^2 P\, ,
\ee
while the variation with respect to $\psi$ gives the Poisson equation
\be\label{8n2Phi1}
8\n^2\Phi=-\kappa^2\rho\, .
\ee
The variations with respect to $\lambda$ and $\gamma$ give a combination of  derivatives of these equations. Observe that \eqs{24ddotPhi}{8n2Phi1} can also be obtained taking directly the variation of the action
(\ref{S2Sintmzro}) with respect to $\Phi$ and $\Psi$, respectively.

The variations with respect to $\beta_i$ (or, equivalently, with respect to $\Xi^i$) and to
 $h_{ij}^{\rm TT}$ give, respectively
\bees
2\n^2\Xi_i &=&-\kappa^2 S_i\label{8n2Phi2}\, ,\\
2\Box h_{ij}^{\rm TT}&=&-\kappa^2\sigma_{ij}\label{8n2Phi3}\, ,
\ees
while the equation obtained performing the variation with respect to $\eps_i$ is  implied by \eq{8n2Phi2}.
Observe that $\Psi$ enters linearly in the action, so it is a Lagrange multiplier. Integrating by parts the term 
$-8\pa_i\Phi\pa^i\Psi$ in the action, the part of the Lagrangian that depends on $\Psi$ is $[(8/\kappa^2)\n^2\Phi+\rho]\Psi$ and the variation with respect to $\Psi$ enforces the constraint (\ref{8n2Phi1}).

Plugging \eq{8n2Phi1} into \eq{24ddotPhi} we can rewrite the latter as
\be\label{ddotPhi}
\ddot{\Phi}+\frac{1}{3}\n^2\Psi =-\frac{4\pi G}{3}(\rho+3 P)\, .
\ee
The term $\ddot{\Phi}$ can be eliminated observing that
 \eq{8n2Phi1} implies that $8\n^2\ddot{\Phi}=-\kappa^2\ddot{\rho}$. Using
\eq{comdT1}, this becomes $8\n^2\ddot{\Phi}=-\kappa^2\n^2\dot{S}$. 
In flat space it is natural to impose 
the boundary condition that $\ddot{\Phi}$ and $\dot{S}$ vanish at infinity. Using again the fact that
a Poisson equation $\n^2f=0$ with $f=0$ at infinity  only  has the solution $f(\vx)=0$, we get
\be\label{ddotPhidotS1}
8\ddot{\Phi}=-\kappa^2\dot{S}\, .
\ee
Therefore \eq{24ddotPhi} can be rewritten as
\be\label{8n2Phi4}
\n^2\Psi =-4\pi G (\rho+3P-3\dot{S})\, ,
\ee
which can be further simplified using \eq{comdT2} to write
$3P-3\dot{S}=-2\n^2\sigma$. 
In conclusion, in Minkowski space we have
\bees
\n^2\Phi &=&-4\pi G\rho\, ,\\
\n^2\Psi &=&-4\pi G (\rho-2\n^2\sigma)\, ,\\
\n^2\Xi_i&=&-16\pi G S_i\, ,\\
\Box h_{ij}^{\rm TT}&=&-16\pi G \sigma_{ij}\label{hijsij}\, .
\ees
We see that  only the tensor perturbations obey a wave equation. The gauge-invariant scalar and vector perturbations obey a Poisson equation, and therefore represent physical but non-radiative degrees of freedom, which are fully determined by the matter distribution. We further observe that
\be
\n^2(\Phi-\Psi)=-8\pi G\n^2\sigma\, .
\ee
Therefore, if the scalar part of the anisotropic stress tensor vanishes, we have $\Phi=\Psi$. 
In the absence of matter we have $\Phi=\Psi=0$ and $\Xi_i=0$, and we only remain with the two radiative degrees of freedom described by $h_{ij}^{\rm TT}$, i.e. with the two polarizations of the massless graviton.

\subsubsection{Lorentz invariance in harmonic variables}\label{sect:Lortransfor}

As with any choice of variables in a theory with  gauge invariance, the use of  the set of harmonic variables $\{\psi, \phi,\gamma, \lambda$, $\beta^i, \eps^i,h_{ij}^{\rm TT}\}$  has some advantages and some drawbacks. The main advantage is that, out of them, we can construct quantities which are invariant under linearized gauge transformations. Furthermore, under spatial rotations the transformations of the harmonic variables are simple: $\psi,\lambda,\gamma$ and $\phi$ are scalar, $\beta^i$ and $\eps^i$ are vectors and
$\hTTij$ is a tensor. In contrast, the behavior of these  variables under Lorentz boosts is  quite complicated (and, as we will see below, even non-local).  For instance, we know that 
the action (\ref{S2Sintmzro}) is Lorentz invariant, since it is just a rewriting of \eq{S2} with different variables. However, in the form (\ref{S2Sintmzro}) 
the Lorentz invariance of the theory is not at all evident.  Furthermore, the scalar, vector and tensor sectors are not separately Lorentz invariant, since of course Lorentz boosts mix scalars with vectors, while vectors mix with scalars and tensors, and tensors mix with  vectors. It is therefore interesting to study in some detail how the harmonic variables transform under Lorentz boosts, and how the invariance of the the action (\ref{S2Sintmzro}) comes out in this formalism. 

A Lorentz transformation of a tensor can be decomposed in two parts, the spin part mixing the Lorentz indices $\mu$ and the orbital part mixing the field indices $x^{\mu}$. It is the spin part that will mix the harmonic variables among themselves since we have decomposed the Lorentz indices $\mu \to (0, i)$. From the point of view of the orbital part, these are just ten  field representations of the Lorentz transformations. This is to make clear that the formalism will lose the manifest Lorentz covariance only as far as the mixing of the tensor components is concerned. 

So consider an infinitesimal  Lorentz transformation $\xm\ra {x'}^{\mu}={\Lambda^{\mu}}_{\nu}\xn$ with ${\Lambda^{\mu}}_{\nu}=\d^{\mu}_{\nu}
+{\omega^{\mu}}_{\nu}$. We then compute the variation $\delta\hmn$ defined as
\be\label{Lorentzhmn}
\delta\hmn\equiv \hmn'(x')-\hmn (x)=
{\omega_{\mu}}^{\rho}\hrn+
{\omega_{\nu}}^{\rho}\hmr\, ,
\ee
so that the orbital part does not appear since we are only interested in how the components mix. Using the expression of the harmonic variables in terms of the metric found in 
sect.~\ref{sect:hardeco} we can  obtain
their variation under Lorentz transformation. We restrict to boosts, i.e. $\omega_{ij}=0$, ${\omega_{0}}^{i}\neq 0$. For $\psi$ and $\phi$ we get 
\bees\label{dpsidphi}
\delta\psi &=&{\omega_{0}}^{i}(\beta_i+\pa_i\gamma)\, ,\\
3\delta\phi &=&-{\omega_{0}}^{i}(\beta_i+\pa_i\gamma)\, .
\label{3deltaphi}
\ees
Observe that $\psi+3\phi$ is Lorentz invariant, as it should since $h=\eMN\hmn =-2(\psi+3\phi)$. Furthermore, these transformations are local functionals of the metric,
since $(\beta_i+\pa_i\gamma)$ is simply $h_{0i}$.
The transformations of $\gamma$ and $\n^2\lambda$ are more complicated and  non-local, since their expression in terms of the metric involves the inverse Laplacian. 
Rather than dealing with the transformation under boosts of the inverse Laplacian it is convenient to start from
$\n^2\gamma=\pa^ih_{0i}$, and observe that, if under a Lorentz transformation a quantity $f$ has a variation $\d f$, then 
\be
\d(\pam f) ={\omega_{\mu}}^{\nu}\pan f +\pam\d f\, , 
\ee
i.e. we treat formally $\pam$  as any other four-vector, satisfying $\d(\pam)={\omega_{\mu}}^{\nu}\pan$. In particular, under boosts 
\be
\d(\pa_0)=\omega_{0i}\pa_i\, ,\qquad
\d(\pa_i)=\omega_{0i}\pa_0\, ,
\ee
and therefore 
\be
\d(\n^2 f)=2{\omega_{0}}^i\pa_i\dot{f}+\n^2(\d f)\, .
\ee 
Thus
$\d(\n^2\gamma)=2{\omega_{0}}^i\pa_i\dot{\gamma}+\n^2(\delta\gamma)$,
while
$\d(\pa^ih_{0i})={\omega_{0}}^i\dot{h}_{0i}+\pa^i\d h_{0i}$, and therefore
\be
\n^2(\delta\gamma)=-2{\omega_{0}}^i\pa_i\dot{\gamma}+
{\omega_{0}}^i\dot{h}_{0i}+\pa^i\d h_{0i}\, .
\ee
This gives
\bees
&&\d\gamma={\omega_{0}}^{i}\n^{-2}\[
\pa_i\(2\psi-\dot{\gamma}-2\phi+\frac{2}{3}\n^2\lambda\) \right. \nn \\ && \hspace*{30mm}\left.+
\(\dot{\beta}_i+\frac{1}{2}\n^2\eps_i\)\]\, .
\ees
Using \eq{n2la} and proceeding similarly, we get
\be
\delta(\n^2\lambda)={\omega_{0}}^{i}\( -\beta_i+2\pa_i\gamma+\frac{3}{2}\dot{\eps}_i\)\, ,\label{Lorlambda}
\ee
while in the vector sector, using  (\ref{invbeta}) and (\ref{inveps}),
we find 
\bees
\delta\beta_i&=&  -{\omega_{0}}^{j} \n^{-2}\[ \pa_i\pa_j (2\psi-2\phi-\dot{\gamma})+\pa_i\dot{\beta}_j\] \nn \\ 
&&+{\omega_{0}}^{j} \left[ (2\psi-2\phi-\dot{\gamma})\d_{ij}
+ \frac{1}{3} \(\pa_i\pa_j-\d_{ij}\n^2\)\lambda \right. \nn\\
 &&\hspace*{9mm}\left. 
 +\frac{1}{2}\pa_j\eps_i
 +h_{ij}^{\rm TT} \right] \, .
 \ees
 \bees\label{epsLorA}
\delta\eps_i=2{\omega_{0}}^j\,\n^{-2}\[ (\pa_i\pa_j-\d_{ij}\n^2)(\dot{\lambda}-\gamma) \right. \nn\\
 \left. 
-\frac{1}{2}\(\pa_i\dot{\eps}_j+\pa_j\dot{\eps}_i\)  +\pa_j\beta_i+\dot{h}_{ij}^{\rm TT}\]\, .
\ees
For the gauge-invariant combinations we get
\bees
\delta\Phi &=&\frac{1}{2}{\omega_{0}}^{i}\Xi_i\, ,\label{deltaPhi}\\
\d\Psi&=&-2{{\omega_0}^i}\n^{-2}\pa_i(\dot{\Phi}+\dot{\Psi})+
{{\omega_0}^i}\Xi_i-\frac{3}{2}{{\omega_0}^i}\n^{-2}\ddot{\Xi}_i\, , \nn\\ \\
\d\Xi_i&=&{\omega_{0}}^{j} \n^{-2}\Bigl[ \Box\hTTij -(\pa_i\dot{\Xi}_j+\pa_j\dot{\Xi}_i) \nn\\ 
&&
\hspace*{14mm} -2(\pa_i\pa_j-\d_{ij}\n^2)(\Phi+\Psi)\Bigr]\, .\\
\d\hTTij&=&\omega_{0i}\Xi_j+\omega_{0j}\Xi_i -\d_{ij}{{\omega_0}^k}\Xi_k\nn\\
&&+{{\omega_0}^k}\n^{-2}\Bigl[\pa_i\pa_j\Xi_k-\pa_i\pa_k\Xi_j-\pa_j\pa_k\Xi_i    \nn\\ 
&&\hspace*{15mm}
-(\pa_i\dot{h}_{jk}^{\rm TT}+\pa_j\dot{h}_{ik}^{\rm TT})\Bigr]\, .
\label{varhTTij}
\ees
Observe that ${\beta'}^i=\beta^i+\d\beta^i$ is transverse with respect to the transformed coordinate ${x'}^i$, i.e. $\pa{\beta'}^i/\pa{x'}^i=0$ or, equivalently,
$\d(\pa_i\beta^i)=0$ (and similarly for $\eps_i$, $\Xi_i$ and $\hTTij$). 

\Eqst{deltaPhi}{varhTTij} show that
under boosts the gauge-invariant variables transform among themselves, although their transformation involves the inverse Laplacian.\footnote{Since under rotations
$\d\Phi=\d\Psi=0$, $\d\Xi_i={\omega_{i}}^{j}\Xi_j$ and 
$\delta\hTTij={\omega_{i}}^{k}h_{kj}^{\rm TT}+
{\omega_{j}}^{k}h_{ki}^{\rm TT}$, these variables transform among themselves under the full Lorentz group.
Of course, one should not be surprised to find six objects that transform among themselves under the  Lorentz group, despite the fact that symmetric traceless tensors form a 9-dimensional irreducible representation of the Lorentz group. The point is that these are representations on fields, and therefore are infinite-dimensional. This is completely analogous to the fact that a transverse vector field $\beta^i(x)$ satisfying $\pa_i\beta^i=0$ is a representation of the rotation  group, despite the fact that it has only two independent (field) components. In other words, a transverse vector field is a  representation of $SO(3)$ of dimension $(\infty)^2$, despite the fact that  a vector is an irreducible  representation of dimension 3.} It is instructive to check explicitly that the action (\ref{S2Sintmzro}) is indeed Lorentz invariant. We neglect for simplicity the interaction term and we split the various terms in \eq{S2Sintmzro} into the scalar, vector and tensor sectors,
\bees
{\cal L}_{2,\rm scalar}&= &-12\dot{\Phi}^2+4\pa_i\Phi\pa^i\Phi
-8\pa_i\Phi\pa^i\Psi\, ,\\
{\cal L}_{2,\rm vector}&= &\pa_i\Xi_j\pa^i\Xi^j\, ,\\
{\cal L}_{2,\rm tensor}&= &
-\frac{1}{2}\pam h_{ij}^{\rm TT}\paM h^{ij,\rm TT}\, .
\ees
Under spatial rotations ${\cal L}_{2,\rm scalar}$, ${\cal L}_{2,\rm vector}$ and
${\cal L}_{2,\rm tensor}$ are of course separately invariant. Under boosts,
using \eqst{deltaPhi}{varhTTij} (and neglecting total derivatives) we get
\bees
\hspace*{-2mm}
\d {\cal L}_{2,\rm scalar}&=&+4{{\omega_0}^i}\Xi_i\n^2(\Phi+\Psi)\, ,\\
\hspace*{-2mm}\d {\cal L}_{2,\rm vector}&=&-4{{\omega_0}^i}\Xi_i\n^2(\Phi+\Psi)
-2{{\omega_0}^i}\Xi^j\Box\hTTij,\\
\hspace*{-2mm}\d{\cal L}_{2,\rm tensor}&=& +2{{\omega_0}^i}\Xi^j\Box\hTTij\, ,
\ees
so the total action is indeed invariant. In contrast ${\cal L}_{2,\rm scalar}$, ${\cal L}_{2,\rm vector}$ and
${\cal L}_{2,\rm tensor}$ are not separately invariant (unless we impose the equations of motion for $\Phi$, $\Psi$ and $\hTTij$,  which in the case $\Tmn=0$ that we are considering read $\n^2\Phi=\n^2\Psi=\Box\hTTij=0$).

\section{Massive gravity in harmonic variables}\label{sect:generic}

We next use these variables to discuss the massive theory linearized over Minkowski space. It will be instructive to  work with a generic Lorentz-invariant mass term, rather than specializing to the FP combination from the beginning. 
It is also convenient to expand
$\gmn =\emn+\kappa\hmn$,
where $\kappa \equiv (32\pi G)^{1/2}$, i.e. to
replace
$\hmn\ra\kappa\hmn$ in the formulas of the previous section, so that henceforth $\hmn$ has  canonical dimensions of mass.\footnote{Thus,  henceforth   $\phi$, $\psi$, $\Phi$ and $\Psi$ have dimensions of mass, $\gamma_i$ is dimensionless and $\lambda$ has dimensions of $[{\rm mass}]^{-1}$ (while, when we perform this decomposition on the dimensionless metric, $\phi$, $\psi$, $\Phi$ and $\Psi$  are dimensionless, $\gamma_i= [{\rm mass}]^{-1}$ and $\lambda= [{\rm mass}]^{-2}$).}
We therefore consider the Lagrangian density
\be\label{SmassiveG2}
{\cal L}_2=\frac{1}{2} \[ 
\hmn {\cal E}^{\mu\nu,\rho\sigma}\hrs
- m_g^2 (b_1 \hmn\hMN+b_2h^2)  \]\, .
\ee
The Pauli-Fierz point corresponds to $b_1=1$, $b_2=-1$. The equations of motion in the absence of matter are now
\be\label{1eqlinmassEmass}
{{\cal E}^{\mu\nu,}}_{\rho\sigma}\hRS
=m_g^2 (b_1\hMN +b_2\eMN h)\, .
\ee
Since $\pan ({{\cal E}^{\mu\nu,}}_{\rho\sigma}\hRS)=0$, they imply the conditions
\be\label{condic1c2}
\pan (b_1\hMN +b_2\eMN h)=0\, .
\ee

\subsection{Elimination of the non-dynamical fields. Scalar sector}\label{sect:elimination}

We first  consider  the scalar sector of the theory. We begin by writing  the action in terms of the variables $\phi,\psi,\lambda,\gamma$ defined in \eqst{hijpsi}{hijphilambdaeps}.
We get
\bees
 {\cal L}_{2,\rm scalar}&=&   -12\dot{\Phi}^2
+4\pa_i\Phi\pa^i\Phi -8\pa_i\Psi\pa^i\Phi\label{LmassiveG}
\\
&&\hspace*{-13mm}-\frac{m_g^2}{2} \biggl[ b_1\( 4\psi^2 +12\phi^2 -2\pa_i\gamma\pa^i\gamma 
+\frac{2}{3} (\n^2\lambda)^2 \)  \nn \\  
&&\hspace*{-2mm}+4b_2 (\psi+3\phi)^2\biggr]\, .\nn
\ees
Since the mass term breaks gauge invariance, the Lagrangian now  depends
on  the four fields $\phi,\psi,\lambda,\gamma$, rather than
just on the two gauge-invariant combinations  $\Phi$ and $\Psi$.  We now want to eliminate the non-dynamical fields and identify the variables which describe the radiative degrees of freedom in the scalar 
sector.   A non-dynamical variable is integrated out by using its {\em  own} equation of motion if it enters the action quadratically, or by using a constraint imposed by a Lagrange multiplier. Such an elimination procedure  goes through even at the path-integral level.
The first step is therefore to chose a convenient set of independent variables.\footnote{Observe that not any choice of variables is legitimate. In particular, the initial conditions on the metric $\hij$ and on its time derivative $\dot{h}_{ij}$ must be in bijection with the initial conditions on the new set of variables. For instance this is  the case for the harmonic variables, but it is not the case for combinations of variables 
 involving e.g. $\dot{\lambda}$ or $\ddot{\lambda}$, such as $\Psi$.
 This point will be important when we consider massive gravity linearized over de~Sitter, in sect~\ref{sect:DS}.\label{foot:change}}

\subsubsection{Outside the FP point: $b_1+b_2\neq 0$.}

We find convenient to  use 
$\Phi,\psi,\lambda,\gamma$  as independent fields. Observe that the change of variables from $\phi$ to $\Phi$ does not involve time derivatives and is thus legitimate, see footnote~\ref{foot:change}.
We discuss separately 
the case  $b_1+b_2\neq 0$ and the case $b_1+b_2=0$ (which includes the FP point $b_1=1, b_2=-1$ and its sign-reversed), since the structure of the non-dynamical equations is different. We begin with the case $b_1+b_2\neq 0$.

By inspection of the Lagrangian (\ref{LmassiveG}) one immediately sees that $\gamma$ and $\psi$ are non-dynamical
\cite{Rubakov:2008nh}. Indeed,
taking the variation  with respect to $\gamma$ (and recalling from \eq{Bard2} that $\Psi$ in \eq{LmassiveG} contains $\dot{\gamma}$, since
$\Psi =\psi -\dot{\gamma}+(1/2)\ddot{\lambda}$) gives
\be\label{Phig1bb}
\n^2(4\dot{\Phi}-b_1 m_g^2\gamma) =0\, . 
\ee
With the boundary condition that $\Phi$ and $\gamma$ vanish at infinity this equation is equivalent to
\be \label{Phig2bb}
4\dot{\Phi}-b_1 m_g^2\gamma=0\, .
\ee
In the massless case this reduces to $\dot{\Phi}=0$, which is in fact the same as 
\eq{ddotPhidotS1}, since we have set $\Tmn=0$.
When $m_g\neq 0$, we can rather use it to eliminate $\gamma$ from the action,
\be\label{elimgamma}
\gamma = \frac{4}{b_1m_g^2}\dot{\Phi}\, .
\ee
Thus, in the massive case $\gamma$ is a non-dynamical variable that can be eliminated algebraically. 
The variation with respect to $\psi$ gives  another algebraic equation,
\be\label{c1c2psi}
4m_g^2(b_1+b_2)\psi =8\n^2\Phi +12m_g^2 b_2\Phi +2m_g^2 b_2\n^2\lambda\, .
\ee
Here we see that the Pauli-Fierz mass term, for which $b_1+b_2=0$, is special. When  $b_1+b_2\neq 0$ we can use \eq{c1c2psi} to eliminate even $\psi$ from the Lagrangian, remaining with a Lagrangian ${\cal L}_2(\Phi,\lambda)$. As expected, we could eliminate two non-physical fields and remain with two degrees of freedom in the scalar sector.

Actually, even $\lambda$ could be integrated out. In fact, taking 
the variation with respect to $\n^2\lambda$ gives
\be\label{n2lambda}
\frac{1}{4}m_g^2(b_1+b_2)\n^2\lambda =\ddot{\Phi}
-\frac{1}{2}m_g^2 (b_1+3b_2)\Phi +\frac{1}{2}m_g^2 b_2\psi\, .
\ee
Thus, for $b_1+b_2\neq 0$ we can eliminate  $\n^2\lambda$ in favor of $\Phi$ and $\ddot{\Phi}$ using \eq{n2lambda} (with $\psi$ expressed in terms of $\n^2\lambda$ and $\Phi$ through \eq{c1c2psi}).
This  results in a Lagrangian that depends only on $\Phi$. However  such a Lagrangian  involves higher-derivative terms proportional to $(\ddot{\Phi})^2$, arising from the terms
$(\n^2\lambda)^2$ in \eq{LmassiveG}. 
A theory whose equations of motion are fourth-order in the time derivatives propagates twice the number of degrees of freedom than  is apparent from  its field content. This is due to the fact that  to evolve the classical equations of motion we need to specify twice as much initial conditions \cite{Hinterbichler:2011tt,Urries}. Furthermore, one of these two degrees of freedom is a ghost \cite{Creminelli:2005qk}.\footnote{Adding to the mass term higher-order polynomials in $\hmn$ with appropriately chosen  coefficients one can have a well-defined Cauchy problem with second-order equations and no ghost, see
\cite{deRham:2010ik,deRham:2010kj,deRham:2011qq}. This is indeed at the basis of the recently proposed ghost-free theory of massive gravity. Here however we are  studying the theory with just a  mass term.}
Thus, if we  eliminate $\n^2\lambda$,  the corresponding (ghost-like) degree of freedom does not disappear, but just hides in the fourth-order equations of motion for $\Phi$. 

For the moment, rather than eliminating $\n^2\lambda$, 
we trade it for a new field $\Gamma$ defined  by 
\be\label{defGamma}
\Gamma =\frac{2}{m_g^2b_2}\( 
-8\n^2\Phi -12m_g^2 b_2\Phi -2m_g^2 b_2\n^2\lambda\)\, ,
\ee
and we use $(\Phi,\Gamma)$ as  independent variables. The variable  $\Gamma$ is defined so that  \eq{c1c2psi} reads
\be\label{elimpsi}
\psi = -\frac{b_2}{8(b_1+b_2)}\, \Gamma\, .
\ee
Eliminating  $\gamma$ and $\psi$  with the help of \eqs{elimgamma}{elimpsi} and expressing $\n^2\lambda$ in terms of $\Gamma$ and $\Phi$ we finally get
\bees \label{PhiGammafull}
{\cal L}_{2,\rm scalar} (\Phi,\Gamma) =
12\dot{\Phi}^2+4\frac{(b_2+4b_1)}{b_2}\pa_i\Phi\pa^i\Phi -12m_g^2b_1\Phi^2\nn\\
+\dot{\Gamma}\dot{\Phi}-
\frac{16(b_1+b_2)}{b_1b_2m_g^2}\pa_i\dot{\Phi}\pa^i\dot{\Phi} 
-\frac{b_1(b_1+2b_2)}{32(b_1+b_2)}m_g^2\Gamma^2 \nn\\ -m_g^2b_1\Gamma\Phi 
-\frac{8(b_1+b_2)}{b_2^2m_g^2}\, (\n^2\Phi)^2+
\frac{(b_1+b_2)}{b_2}\,\pa_i\Gamma\pa^i\Phi\, .\nn \\
\ees
With a standard analysis, discussed in App.~\ref{app:A}, we can recover from this Lagrangian the known result that one of these two degrees of freedom is a ghost while the other has the correct sign for the kinetic term. In particular, for $(b_1+b_2)/b_1b_2\leq 0$, the ``healthy" mode is $\Phi$ while the ghost is $\Gamma$.\footnote{For 
\label{foot:ghost}
$(b_1+b_2)/b_1b_2> 0$ the situation is  more involved. Because of the term $\pa_i\dot{\Phi}\pa^i\dot{\Phi}$ in \eq{PhiGammafull},   as
we discuss in App.~\ref{app:A}
there is a critical value 
$k_*\equiv |\vk_*|$ defined by 
$({k_*}/{m_g})^2=(3/4)\, {(b_1b_2)}/{(b_1+b_2)}$,
such that are ghost-like the  Fourier modes 
$\Gamma_{\bf k}$ with $|\vk|<k_*$ and the Fourier modes
$\a_k\Phi_{\bf k}+\Gamma_{\bf k}$ with $|\vk|>k_*$, where $\a_k$ is 
defined in \eq{AkBk}. However, even in this case, as $b_1+b_2\ra 0$ we have  $k_*\ra+\infty$ and
only the Fourier modes of $\Gamma$ are ghost-like.
} 
Concerning the masses of the two scalar modes, as we show in
App.~{\ref{app:A},
the poles of the propagator are at 
\be\label{mssPhi}
m^2=\left\{ b_1 m_g^2,\,\, \frac{b_1 (b_1+4b_2)}{2 (b_1+b_2)} m_g^2\right\}\, .
\ee
In particular, setting $b_1=1, b_2=-1-\eps$ the second pole, which corresponds to the ghost, is at 
\be\label{massGamma}
m^2 =\frac{3+4\eps}{2\eps} m_g^2\, .
\ee

\subsubsection{Approaching the FP point,  $b_1+b_2\ra 0$ with $b_1>0$}

\Eq{massGamma} shows that the ghost mass diverges as $\eps\ra 0$. In this limit we can also easily see this  directly from the Lagrangian,   writing
$b_1=1$ and $b_2=-1-\eps$, and neglecting terms that vanish as $\eps\ra 0$. Then 
\eq {PhiGammafull} becomes
\bees\label{PhiGammafulleps}
{\cal L}_{2,\rm scalar} (\Phi,\Gamma) &=&12\[ \dot{\Phi}^2
-\pa_i\Phi\pa^i\Phi-m_g^2\Phi^2\]
+\dot{\Gamma}\dot{\Phi} \nn\\ && -\frac{1+2\eps}{32\eps}m_g^2\Gamma^2
-m_g^2\Gamma\Phi \, .
\ees
Note also  that the term proportional to $\pa_i\dot{\Phi}\pa^i\dot{\Phi}$ disappeared, and therefore the ghost is simply $\Gamma$, see footnote~\ref{foot:ghost}.
Diagonalizing the kinetic term as in App.~{\ref{app:A},
\bees
{\cal L}_{2,\rm scalar} (\Phi,\Gamma) &=&\frac{1}{48}| 24\dot{\Phi}+\dot{\Gamma}|^2-
\frac{1}{48}|\dot{\Gamma}|^2
-12\pa_i\Phi\pa^i\Phi \nn\\
&& -12m_g^2\Phi^2-\frac{1+2\eps}{32\eps}m_g^2\Gamma^2
-m_g^2\Gamma\Phi \, .\nn\\
\ees
As $\eps\ra 0$ the mass term for $\Gamma$ diverges,
and $\Gamma$ effectively disappears from the theory. This result is independent of whether we take the limit $\eps\ra 0^+$ or $\eps\ra 0^-$, i.e. of whether the mass term for $\Gamma$ is tachyonic or not. This can be seen more clearly
observing that the equation of motion  derived from
the Lagrangian density (\ref{PhiGammafulleps}) taking the variation with respect to $\Gamma$ is
\be\label{eqmotG}
(1+2\eps) \frac{m_g^2}{16}\Gamma =-\eps (\ddot{\Phi}+m_g^2\Phi)\, ,
\ee
which in the limit $\eps\ra 0$ gives $\Gamma =0$. 
Therefore at the FP point the ghost  disappears
and, taking the limit $\eps\ra 0$ in \eq{PhiGammafulleps},
we remain with
\bees\label{Phicovariant}
\hspace*{8mm}{\cal L}_{2,\rm scalar}&=&12 
\[\dot{\Phi}^2-\pa_i\Phi\pa^i\Phi-m_g^2\Phi^2\] \nn\\
&=&-12(\pam\Phi\paM\Phi +m_g^2\Phi^2)\, .
\ees
This Lagrangian density has the correct (i.e. non ghost-like) sign for the kinetic term, and the correct non-tachyonic sign for the mass term. The equation of motion for $\Phi$ at the FP point is the massive Klein-Gordon (KG) equation,
\be\label{BoxPhimm}
(\Box-m_g^2)\Phi=0\, .
\ee
Thus the Bardeen ``potential" $\Phi$, that in the massless case is a physical but non-radiative degree of freedom, at the FP point becomes a  radiative field which describes the 
helicity-0 component of the massive graviton. This  is a result that was obtained already in 1966 in \cite{Deser:1966zzb} by performing the elimination of the non-dynamical degrees of freedom in the ADM formalism,\footnote{In \cite{Deser:1966zzb} the scalar sector of the spatial metric was parametrized as
$h_{ij}=(1/2)(\d_{ij}-\n^{-2}\pa_i\pa_i)h^T+2\pa_i\pa_jh^L$ and
it was indeed found that the Hamiltonian has the KG form in terms of the variable $h^T$  and of its conjugate momentum. In our notation
$h^T=-4\phi-(2/3)\n^2\lambda=4\Phi$.
} and more recently in \cite{Alberte:2010it}.

At the technical level it is remarkable that, at the FP point,
the complicated Lagrangian density (\ref{PhiGammafull})
collapses to the above simple KG form, particularly considering that $\Phi$ is {\em not} a Lorentz scalar. 
We have  indeed seen in \eq{deltaPhi} that under boosts it  mixes with $\Xi^i$. In particular, the Lagrangian (\ref{Phicovariant}) is not Lorentz invariant. The Lorentz invariance of the theory only emerges when we recombine the scalar, vector and tensor sectors.
However, the appearance of a KG equation is expected on physical grounds since, once we eliminate all the non-dynamical variables, the remaining physical degree of freedom must satisfy a Lorentz-invariant dispersion relation.

\subsubsection{Working  directly at the FP point $b_1=1, b_2=-1$.}

It is  instructive to derive the above results working directly at the FP point, rather than approaching it as a limit, since the structure of the equations is different.
The variable $\gamma$ can still be eliminated using the variation with respect to $\gamma$, that now becomes
\be\label{elimgammaFP}
\gamma = \frac{4\dot{\Phi}}{m_g^2}\, .
\ee
The equation obtained from the variation with respect to $\psi$  now has  a different structure:
the term proportional to $\psi^2$ in \eq{LmassiveG}  cancels, and $\psi$ enters in the Lagrangian ${\cal L}_{2,\rm scalar}$ only linearly,
\be
{\cal L}_{2,\rm scalar}= {\cal L}'_{2,\rm scalar} +\psi \[ 8\n^2\Phi -12m_g^2 \Phi -2m_g^2 \n^2\lambda\] \, ,
\ee
where ${\cal L}'_{2,\rm scalar}$ is independent of $\psi$. 
Thus $\psi$ is a Lagrange multiplier that imposes the condition
\be\label{varpsiFP}
8\n^2\Phi -12m_g^2 \Phi -2m_g^2 \n^2\lambda
=0\, ,
\ee
which is nothing but  $\Gamma=0$. The variation with respect to $\psi$ therefore eliminates the ghost. 
After eliminating $\gamma$  through \eq{elimgammaFP} and using the condition $\Gamma=0$
 we remain with a Lagrangian function  of $\Phi$ only which, as expected, gives back \eq{Phicovariant}.

\subsubsection{Interaction with matter sources at the FP point}\label{sect:intmat}

We now include the coupling to the energy-momentum tensor, to see how the elimination of the non-dynamical variables works in the presence of matter and what quantity acts as a source for the radiative field $\Phi$. 
From \eq {S2Sintmzro}  the interaction Lagrangian in the scalar sector is
\be
{\cal L}_{\rm int, scalar}=\kappa (3\Phi P+\Psi\rho)\, ,
\ee
where the factor of $\kappa$  comes from the fact that here the harmonic decomposition is performed on the metric perturbation $\hmn$ defined by $\gmn=\emn+\kappa\hmn$, while in Sect.
\ref{sect:harmo} we were using  $\gmn=\emn+\hmn$.
Adding  this interaction term to  the Lagrangian in \eq{LmassiveG}, specializing to the FP point, and writing the Lagrangain in terms of the independent variables 
$\{\gamma, \psi, \lambda,\Phi\}$, we get
\bees
({\cal L}_{2}+{\cal L}_{\rm int})_{\rm scalar}&=&
-12\dot{\Phi}^2+4\pa_i\Phi\pa^i\Phi -8\pa_i\psi\pa^i\Phi \nn\\ 
&&\hspace*{-25mm} +
8\pa_i\dot{\gamma}\pa^i\Phi+4\ddot{\Phi}\n^2\lambda\nn\\
&&\hspace*{-25mm}+m_g^2 \[ 12\Phi^2+ 4\Phi\n^2\lambda +\pa_i\gamma\pa^i\gamma 
-12\Phi\psi -2\psi \n^2\lambda\]\nn\\
&&\hspace*{-25mm}+\frac{\kappa}{2} [ 6\Phi P+ 2\psi\rho-2\dot{\gamma}\rho+\dot{S}\n^2\lambda]\,.
\label{L2+Lint}
\ees
In the last line we used the conservation equation (\ref{comdT1}) and integration by parts
to replace
$\ddot{\lambda}\rho\ra \lambda\ddot{\rho} = \lambda\n^2\dot{S}\ra \dot{S}\n^2\lambda$.
The variations with respect to $\gamma, \psi$ and $\lambda$ give, respectively,
\bees
\n^2(8\dot{\Phi}-2 m_g^2\gamma) &=&-\kappa\dot{\rho}\, . \label{Phig1bbsource}\\
8\n^2\Phi -12m_g^2 \Phi -2m_g^2 \n^2\lambda
&=&-\kappa\rho\, .\label{varpsiFPsource}\\
2\ddot{\Phi}+2m_g^2 \Phi- m_g^2 \psi&=&-\frac{1}{4}\kappa\dot{S}\, . 
\label{elimpsiatFPsource}
\ees
Using again the conservation equation (\ref{comdT1}), $\dot{\rho} =\n^2 S$,  we can rewrite \eq{Phig1bbsource} as
\be
m_g^2\gamma =4\dot{\Phi}+\frac{\kappa}{2}S\, .
\ee
We now eliminate $\gamma$ from the Lagrangian density (\ref{L2+Lint}) using this equation, and
$\n^2\lambda$ and $\psi$ using \eqs{varpsiFPsource}{elimpsiatFPsource}, respectively.
We get ${\cal L}_{\rm scalar}= {\cal L}_{2,\rm scalar}+{\cal L}_{\rm int, scalar}$ with 
\be
{\cal L}_{2,\rm scalar}=
-12(\pam\Phi\paM\Phi +m_g^2\Phi^2) \, .
\ee
and
\be\label{Lintscal}
{\cal L}_{\rm int, scalar}=\kappa\Phi (2\rho+3P-3\dot{S})\, .
\ee
Observe that all terms quadratic in the sources, which appear in the intermediate steps of the computation, in the end canceled.
The equation of motion for $\Phi$ in the presence of sources is therefore
\be\label{KGPhiSource}
(\Box-m_g^2)\Phi=-\frac{\kappa}{24} (2\rho+3P-3\dot{S})\, ,
\hspace{5mm} ([\Phi]\sim {\rm mass}),
\ee
where we have recalled that in this section the variable $\Phi$ was defined from the harmonic decomposition of the field $\hmn$ defined from $\gmn=\emn+\kappa\hmn$, so both $\hmn$ and $\Phi$ have the canonical dimensions of mass. If instead we use the more common definition of $\Phi$  from the harmonic decomposition of the   field $\hmn$ defined from $\gmn=\emn+\hmn$ (as we did in Sect.~\ref{sect:harmo}), then both $\hmn$ and $\Phi$ are dimensionless, and 
\eq{KGPhiSource} becomes
\bees\label{KGPhiSource2}
(\Box-m_g^2)\Phi&=&-\frac{\kappa^2}{24} (2\rho+3P-3\dot{S})\,\quad
 ([\Phi]\sim 1).
\ees
Using \eq{comdT2}, we can also rewrite this as
\be\label{KGPhiSource3}
(\Box-m_g^2)\Phi
=-\frac{8\pi G}{3} (\rho-\n^2\sigma)\, ,
\hspace{10mm} ([\Phi]\sim 1),
\ee
showing that, at the level of linearized theory, the helicity-0 component of the massive graviton is sourced by the combination $(\rho-\n^2\sigma)$. Observe  that in contrast, in the massless case, $\Phi$ was coupled only to $\rho$, and through a Poisson equation.

\subsubsection{The higher-derivative action for $\Phi$}

We now follow a different route and, for $b_1,b_2$ generic, we  eliminate $\Gamma$ and obtain a higher-derivative Lagrangian for $\Phi$. From \eq{PhiGammafull} we see that the variation with respect to $\Gamma$ gives
\be
\Gamma = -\frac{16 (b_1+b_2)}{b_1(b_1+2b_2)m_g^2}\, 
\( \ddot{\Phi}+m_g^2 b_1\Phi +\frac{b_1+b_2}{b_2}\n^2\Phi\)\, .
\ee
Thus, as we already remarked, even $\Gamma$ can be eliminated algebraically, but the price to be paid is a higher-derivative action for $\Phi$. We plug this expression for $\Gamma$ into \eq{PhiGammafull} and, after some integration by parts,  we find that all terms  ``miraculously" combine to form explicitly Lorentz-covariant structures,
\be\label{miracle}
{\cal L}_2(\Phi)=\a_1 \pam\Phi\paM\Phi +
\a_2\,m_g^2\Phi^2 +\frac{\a_3}{m_g^2}\, 
(\Box\Phi)^2\, ,
\ee
with
\be\label{a1a2b}
\a_1=\frac{4(b_1-2b_2)}{b_1+2b_2}\, ,\quad
\a_2=-\frac{4b_1(b_1+4b_2)}{b_1+2b_2}\, ,
\ee
\be\label{a3b}
\a_3=\frac{8(b_1+b_2)}{b_1(b_1+2b_2)}\, .
\ee
At the FP point the higher-derivative term disappears and the result reduces to \eq{Phicovariant}, as it should. 
The term $(1/m_g^2)(\Box\Phi)^2$ shows that, without the FP tuning, the theory becomes strongly coupled already at the scale $m_g$. This is a result which is usually derived  introducing a scalar  {\Stu} field, but which is quite transparent also in our formalism.

Including also the interaction with the energy-momentum tensor, and repeating for $b_1,b_2$ generic the computations done at the FP point in sect.~\ref{sect:intmat}, after long but straightforward algebra we get 
\bees\label{miracle2}
{\cal L}_{\rm int}(\Phi) &=&\kappa\Phi \( 
\frac{2b_2}{b_1+2b_2}\rho+3P-\frac{b_1+4b_2}{b_1+2b_2}\dot{S}\) \nn \\
&&+\frac{2(b_1+b_2)\kappa}{b_1(b_1+2b_2)m_g^2} (\rho-\dot{S})\Box\Phi\nn\\
&&+\frac{(b_1+b_2)\kappa^2}{8b_1(b_1+2b_2)m_g^2} (\rho-\dot{S})^2\, .
\ees
As it should, at the FP point the result reduces to \eq{Lintscal}. Observe however that outside the FP point there is also a  term quadratic in the sources, proportional to
$(\rho-\dot{S})^2$.

\subsection{Elimination of the non-dynamical fields. Vector and tensor sector}\label{sect:vect}

A similar analysis can be performed in the vector sector of the theory, see also
\cite{Rubakov:2008nh,Alberte:2010it}. With the addition of a generic mass term the quadratic part of the Lagrangian density in the vector sector reads
\be
{\cal L}_{2,\rm vector}=\pa_i\Xi_j\pa^i\Xi^j+m_g^2b_1
\(\b_i\beta^i-\frac{1}{4}\pa_i\eps_j\pa^i\eps^j\)\, ,
\ee
where $\Xi_i=\beta_i-(1/2)\dot{\eps}_i$.  We use $\beta_i$ and $\eps_i$ as independent fields. We see that $\beta_i$ is a non-dynamical field, and the variation with respect to it gives
\be\label{elimbetai}
2\n^2\beta^i-\n^2\dot{\eps}^{\, i}-2m_g^2b_1\beta^i=0 \, .
\ee
In contrast $\eps_i$ is dynamical, since $\Xi_i$ contains $\dot{\eps}_i$, and its
variation gives
\be\label{eqeps}
\n^2\({\ddot{\eps}}^{\, i}-2\dot{\beta}^i+m_g^2b_1\eps^i\)=0\, .
\ee
With the boundary condition that the fields vanish at infinity, this is equivalent to
\be\label{eqmotepsi}
\ddot{\eps}^{\, i}-2\dot{\beta}^i+m_g^2b_1\eps^i=0\, .
\ee
Taking the time derivative of \eq{elimbetai} and substituting $\ddot{\eps}^{\, i}$ from
\eq{eqmotepsi} we get the relation
\be\label{conditionepsbeta}
\n^2\eps^i=2\dot{\beta}^i\, ,
\ee
which inserted back into \eq{eqmotepsi} gives
\be\label{KGepsi}
(\Box-m_G^2)\eps^i=0\, ,
\ee
where $m_G^2=b_1 m_g^2$.
Thus,  the mass of the two vector modes described by the transverse vector $\eps^i$ is the same as the mass of the scalar mode. Clearly, $\eps^i$ describes the two components of the massive graviton with helicities $\pm 1$.\footnote{We adhere here to a common abuse of language. Of course, these modes become helicity eigenstates only in the massless limit.}

For $b_1<0$, however, these vector modes are not only tachyonic (since
$m_G^2=b_1 m_g^2$ becomes negative) but even  
ghost-like. The simplest way to  understand this point  is to express the Lagrangian in terms of  the momentum modes, 
performing  a spatial  Fourier transform. From \eq{elimbetai}, for
the Fourier modes 
$\beta^i_{\vk}(t)$ and $\eps^i_{\vk}(t)$  we get
\be
\beta^i_{\vk}=\frac{1}{2}\, \frac{\vk^2}{\vk^2+m_g^2b_1}\, \dot{\eps}^{\, i}_{\vk}\, .
\ee
This allows one to eliminate $\beta^i_{\vk}$ from the action. Then, for
the Lagrangian $L=\int d^3x {\cal L}$ one
obtains~\cite{Rubakov:2008nh}
\bees
  L_{2,\rm vector} (\eps_i)&=& 
\frac{m_g^2b_1}{4}\, \int \frac{d^3k}{(2\pi)^3}\, \frac{\vk^2}{\vk^2+b_1m_g^2} \\&&
\hspace*{-5mm}\times
\[ \dot{\eps}_{i,\vk}^*\dot{\eps}^i_{\vk}
-\eps_{i,\vk}(\vk^2+b_1m_g^2)\eps^i_{\vk}\]\, . \nn 
\ees
If $b_1>0$ we have $m_G^2=b_1 m_g^2>0$, and 
\bees\label{Lvect}
L_{2,\rm vector} (\eps_i)&=& \frac{m_G^2}{4}\,\int \frac{d^3k}{(2\pi)^3} \,
 \frac{\vk^2}{\vk^2+m_G^2}\\ &&  \times\, 
\[ \dot{\eps}_{i,\vk}^*\dot{\eps}^i_{\vk}
-\eps_{i,\vk}^*(\vk^2+m_G^2)\eps^i_{\vk}\]\, . \nn 
\ees
We see that the kinetic term has the good, non-ghostlike, sign and the mass term is non-tachyonic. For each given momentum mode $\eps^i_{\vk}$, the equation of motion derived from
the Lagrangian (\ref{Lvect}) is the same as that derived from
a Lagrangian $\[ \dot{\eps}_{i,\vk}^*\dot{\eps}^i_{\vk}
-\eps_{i,\vk}^*(\vk^2+m_G^2)\eps^i_{\vk}\]$, which is just a massive Klein-Gordon Lagrangian, so we recover the massive  KG equation (\ref{KGepsi}).
In contrast, if $b_1<0$ the sign of the kinetic term in the action is such that the modes of
$\eps^i_{\vk}$ with $|\vk|^2>-b_1 m_g^2$ are ghost-like. This eliminates the possibility of a consistent theory at the sign-reversed of the FP point, $b_1=-1, b_2=1$.

Observe also that, going back to coordinate space, the Lagrangian (\ref{Lvect}) 
becomes~\cite{Alberte:2010it}
\bees
 L_{2,\rm vector} (\eps_i)&=&  -\frac{m_G^2}{4}\, 
\int d^3x \,\nn \\ && \hspace*{-10mm}  \times  \frac{\n^2}{\n^2-m_G^2}\, 
\( \pam\eps_i\paM\eps^i+m_G^2\eps_i\eps^i\)\, .
\ees
The non-locality of this expression with respect to the spatial coordinates is ultimately a consequence of the fact that $\eps^i$ is a non-local function of the metric. However we have seen above that in the end,   the equation of motion for $\eps^i$ is just a massive KG equation, so it is local.

We next add the interaction term. Using \eq{S2Sintmzro} we see that in the vector sector
(with the present normalization of the fields) it is given  by 
${\cal L}_{\rm int}=-\kappa\beta_iS^i +(\kappa/2)\dot{\eps}_iS^i$. Then
\eqs{elimbetai} {eqeps} become
\bees
2\n^2\beta^i-\n^2\dot{\eps}^{\, i}-2m_G^2\beta^i&=&-\kappa S^i
\label{elimbetai2} \, ,\\
\n^2\({\ddot{\eps}}^{\, i}-2\dot{\beta}^i+m_G^2\eps^i\)&=&\kappa \dot{S}^i\, .
\label{eqeps2}
\ees
Using the conservation equation (\ref{comdT3}) and the usual boundary conditions that the fields vanish at infinity, \eq{eqeps2} becomes
\be\label{eqeps3}
{\ddot{\eps}}^{\, i}-2\dot{\beta}^i+m_G^2\eps^i= \frac{\kappa}{2} \sigma^i\, .
\ee
As in the sourceless case, taking the time derivative of \eq{elimbetai2} and combining it with \eq{eqeps2} gives $\n^2\eps^i=2\dot{\beta}^i$. Plugging this into \eq{eqeps3} we  get
\be\label{KGepsisource}
(\Box-m_G^2)\eps^i=-\frac{\kappa}{2} \sigma^i\, .
\ee
Finally, the tensor sector is simple, since we only have $\hTTij$. Using
\eq{S2Sintmzro},
the Lagrangian density in the tensor sector is
\bees
&&({\cal L}_2+{\cal L}_{\rm int})_{\rm tensor}=\\
&&-\frac{1}{2} \[ 
\parho h_{ij}^{\rm TT}\paR h^{ij,\rm TT}
+ m_g^2 b_1 h_{ij}^{\rm TT} h^{ij,\rm TT} \] 
+\frac{\kappa}{2}h_{ij}^{\rm TT}\sigma^{ij}\, .\nn
\ees
Again, $b_1>0$ gives a non-tachyonic mass term, while for $b_1<0$ we have a tachyon.
The equation of motion is
\be
(\Box-m_G^2)h^{ij,\rm TT}=-\frac{\kappa}{2} \sigma^{ij}\, ,
\ee
Comparing with \eq{mssPhi} we see that
even when $b_1\neq 1$  the scalar $\Phi$, the vector $\eps^i$ and tensor modes 
$h^{ij,\rm TT}$, that together make up the five components of the massive graviton,  have the same mass,  given by $m_G^2=b_1m_g^2$. This is in agreement with the representation theory of the Poincar\'e group, as it should for a theory  linearized over flat space.

\section{A hidden gauge symmetry}\label{sect:hidden}

The results of the previous section show that, out of the 10 components of the metric,
the four variables $\gamma$, $\psi$ and $\beta^i$ (where $\beta^i$ is transverse and therefore carries two degrees of freedom) are non-physical and can be eliminated through their equations of motion. The five  components of the massive graviton are described by
$\{\Phi, \eps_i,h_{ij}^{\rm TT}\}$, with $\eps^i$ transverse and $h_{ij}^{\rm TT}$ transverse-traceless. Finally, the ghost is given by a linear combination of $\n^2\lambda$ and $\Phi$, and disappears at the FP point. In particular, when $(b_1+b_2)/b_1b_2\leq 0$, the ghost is given by the combination $\Gamma$ defined in \eq{defGamma}. Alternatively, both scalar degrees of freedom can be described by the field $\Phi$, through a higher-derivative action.

We can describe this as a reduction process from the original action
$S[\hmn]$, which is a functional of all the 10 components of the metric, to a reduced 
action  $S_{\rm red}[\Phi, \eps_i,h_{ij}^{\rm TT}]$ which depends only on the physical degrees of freedom (and includes a higher-derivative term proportional
$(\Box\Phi)^2$  if we are not at the FP point),
\be
S[\hmn]\longrightarrow
S_{\rm red}[\Phi, \eps_i,h_{ij}^{\rm TT}]\, .
\ee
This result, however, brings a surprise. Because of  the mass term, the starting action  $S[\hmn]$ is not gauge invariant. Nevertheless,  the variables $\Phi$ and $h^{ij,\rm TT}$ are gauge invariant, while $\eps^i$  transforms as $\eps\ra \eps_i-2B_i$,  see \eq{tr3}. 
Therefore, 
none of these  fields is  sensitive to the transformations parametrized by $A$ and $C$, so the reduced action is trivially   invariant under diffeomorphisms of  the form 
\be\label{xired}
\xi_0=A\, ,\qquad \xi_i=\pa_i C\, .
\ee
Observe that this is true independently of the form of the mass term. More precisely, if we perform a gauge transformation (\ref{xired}), the original  action
$S[\hmn]$ transforms into a different quantity $S'[\hmn;A,C]$. However, as illustrated graphically
in fig.~\ref{fig:fig1},
the reduction to the physical degrees of freedom performed with $S'[\hmn;A,C]$ (using of course its own non-dynamical equations of motion) gives the same reduced action as  that obtained from $S[\hmn]$ 
\be
S'[\hmn;A,C]\longrightarrow
S_{\rm red}[\Phi, \eps_i,h_{ij}^{\rm TT}]\, ,
\ee
independently of $A$ and $C$. We therefore have a ``hidden" symmetry at the level of the original action: even if $S[\hmn]$ is not invariant under a gauge transformation of the form (\ref{xired}), still it is transformed to an action 
$S'[\hmn;A,C]$ that describes the same physics. In this sense, \eq{xired} is a symmetry of the original action $S[\hmn]$.\footnote{To avoid misunderstanding, observe that we cannot use $A$ and $C$ to fix a gauge in the original theory, e.g. setting $\lambda =\gamma=0$ 
in the equations of motion, as we could do in the massless theory. 
In fact, using the Lagrangian ${\cal L}$, the variables $\lambda$ and $\gamma$ are fixed in terms of $\Phi$ 
by \eqs{elimgammaFP}{varpsiFP} (or by their generalizations outside the FP point), and cannot be set to zero.
Using the Lagrangian ${\cal L}'$ they are instead given by 
\eqs{elimgammaFPC}{varpsiFPC} below. In this case one might choose $\n^2C$ and $A$ such that $\lambda =\gamma=0$. However, the resulting functions $A$ and $C$ would now depend on $\Phi$, and would contribute to the action (\ref{DeltaLAC}). Thus, the symmetry that we are discussing is different from a usual gauge symmetry, in which the action is invariant under the gauge transformation and the gauge freedom  can be used to remove some degrees of freedom from the theory. In our case the original action is not invariant, but changes to a new action $S'[\hmn;A,C]$. The only sense in which these transformations are a symmetry of the theory is the one illustrated graphically in fig.~\ref{fig:fig1}.}

\begin{figure}
\includegraphics[width=0.45\textwidth]{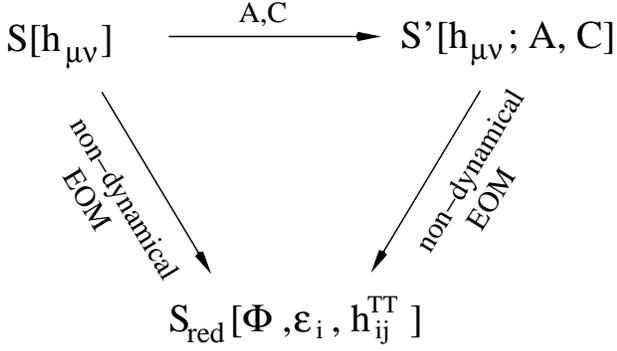}
\caption{\label{fig:fig1} A graphic representation of the relations between $S$, $S'$ and the reduced action.}
\end{figure}

To illustrate this equivalence  we consider 
the  original action  in the scalar sector, \eq{LmassiveG}, and we study how it changes under a transformation parametrized by $A$ and $C$. We work with  $b_1,b_2$ generic and we use  $\{\Phi,\psi,\gamma, \lambda\}$ as independent fields. 
We then transform these variables according to \eqst{tr1}{tr3}
and we find that the transformed  Lagrangian density  is  
\bees
&&{\cal L}'_{2,\rm scalar}[\Phi,\psi,\gamma,\lambda;A,C]=  \\
&&{\cal L}_{2,\rm scalar}[\Phi,\psi,\gamma,\lambda] +
\Delta{\cal L}_{2,\rm scalar}[\Phi,\psi,\gamma,\lambda;A,C]\, ,\nn 
\ees
where ${\cal L}_{2,\rm scalar}$ is the original Lagrangian. Written in terms of the variables $\{\Phi,\psi,\gamma, \lambda,\}$  it reads
\bees\label{dSC}
&& {\cal L}_{2,\rm scalar}[\Phi,\psi,\gamma,\lambda]=   \\ &&
 -12\dot{\Phi}^2+4\pa_i\Phi\pa^i\Phi -8\pa_i\psi\pa^i\Phi+
8\pa_i\dot{\gamma}\pa^i\Phi+4\ddot{\Phi}\n^2\lambda\nn\\
&& \hspace*{-3mm}+m_g^2  \Big[ -2(b_1+b_2)\psi^2-6(b_1+3b_2)\Phi^2
-2(b_1+3b_2)\Phi\n^2\lambda\nn\\
&& -\frac{1}{2}(b_1+b_2)(\n^2\lambda)^2
+b_1\pa_i\gamma\pa^i\gamma
+12b_2\Phi\psi +2b_2\psi \n^2\lambda\Big]\, . \nn 
\ees
The extra term is
\bees
&&\hspace*{-4mm}\Delta{\cal L}_{2,\rm scalar}[\Phi,\psi,\gamma,\lambda;A,C]= \\ &&\hspace*{-4mm} m_g^2 \dot{A}\[ 4(b_1+b_2)\psi-12b_2\Phi-2b_2\n^2\lambda\]
-2b_1m_g^2\pa^iA\pa_i\gamma \nn\\ &&\hspace*{-4mm}+m_g^2 \n^2C \[ 2(b_1+b_2)\n^2\lambda+4(b_1+3b_2)\Phi-2b_1\dot{\gamma}
-4b_2\psi\]\nn\\
&&\hspace*{-4mm}-2(b_1+b_2) m_g^2 (\dot{A}-\n^2C)^2
+b_1m_g^2\pa^i(A-\dot{C})\pa_i(A-\dot{C})
\, . \nn\label{DeltaLAC}
\ees
We can now take the variation of
${\cal L}_{2,\rm scalar}+\Delta{\cal L}_{2,\rm scalar}$ 
with respect to 
$\gamma$ and $\psi$. Specializing for simplicity to the FP point we get, respectively,
\bees
m_g^2\gamma &=& 4\dot{\Phi}+m_g^2(A+\dot{C})\, ,\label{elimgammaFPC}\\
m_g^2 \n^2\lambda&=&  4\n^2\Phi -6m_g^2 \Phi +2m_g^2\n^2C\, ,\label{varpsiFPC}
\ees
Not surprisingly, these equations could have been obtained simply performing the replacements \eqst{tr1}{tr2} directly into  \eqs{elimgammaFP}{varpsiFP}.
Substituting these expressions for $\gamma$ and $\n^2\lambda$  back into \eqs{dSC}{DeltaLAC} (with $b_1=1, b_2=-1$) we find that, in ${\cal L}_{2,\rm scalar}+\Delta{\cal L}_{2,\rm scalar}$,  all terms that depend on $A$ and $C$ (both linear and quadratic) cancel, and therefore the reduced action is the same as that found in the case $A=C=0$. The same calculation can be performed for $b_1+b_2\neq 0$, and we have checked that again the extra terms cancel upon use of the equations of motion that eliminate $\gamma,\psi$ and $\n^2\lambda$ (or, equivalently, $\Gamma$).

The symmetry that we have found takes a simpler form if, rather than using $(A,C)$ as independent gauge functions, we introduce 
\be
\bar{A}\equiv A-\dot{C}\, ,
\ee
and we use the pair $(\bar{A},C)$ as independent variables. Then the residual gauge symmetry  consists of linearized diffeomorphisms with 
\be\label{ximApamC}
\xi_{\mu}=\bar{A}\delta_{\mu}^0  +\pam C\, .
\ee
In particular the transformation parametrized by $C$ takes a nice covariant form,
\be
\hmn\ra\hmn -2\pam\pan C\, .
\ee
It is interesting to understand the origin of this hidden symmetry using this covariant form. Under $\hmn\ra\hmn -2\pam\pan C$ the term 
$\hmn {\cal E}^{\mu\nu,\rho\sigma}\hrs$ in \eq{SmassiveG2} is obviously invariant, and transforming the mass term we get 
\bees\label{c1c2BoxC}
{\cal L}'_2[\hmn;C]=\frac{1}{2} \[ 
\hmn {\cal E}^{\mu\nu,\rho\sigma}\hrs
- m_g^2 (b_1 \hmn\hMN+b_2h^2)  \]\nn\\
-2m_g^2 \[(\pan C)\pam  (b_1\hMN+b_2\eMN h)
+(b_1+ b_2) (\Box C)^2)\]\, . \nn\\
\ees
Observe that, at the FP point, the term quadratic in $C$  drops. This corresponds to the well known fact that, when  $\hmn$ takes the form $\hmn=\pam\pan \phi$ for some function $\phi$, the FP mass term is a total derivative. The generalization of this property to terms of cubic and higher order in $\pam\pan\phi$ gives rise to the Galileon family of operators~\cite{Nicolis:2008in}. 

In this covariant formulation a simple but not totally correct way of showing that, upon use of the non-dynamical equations of motion, the extra term in the Lagrangian vanishes, is as follows. 
Define
\be\label{defjnu}
j^{\nu}\equiv \pam (b_1\hMN +b_2\eMN h)\, .
\ee
We see from \eq{condic1c2} that, if we use the equations of motion of the starting Lagrangian ${\cal L}_2$,  we have $j^{\nu}=0$. However, as we show in app.~\ref{app:jnu}, to derive this condition it is not sufficient to use the non-dynamical equations of motion. Rather, we also need the equation of motion for $\eps^i$, which is a dynamical variable. Still, using only non-dynamical equations of motion, we can show that
\be
j^0=0\, ,\qquad \pa_ij^i=0\, ,
\ee
and therefore also
\be\label{panjn}
\pan j^{\nu}=0\, .
\ee
Thus, as a consequence of the equations that eliminate the non-dynamical degrees of freedom, the term $(\pan C)j^{\nu}$ in \eq{c1c2BoxC} vanishes upon integration by parts.
The term $(\Box C)^2$ is anyhow irrelevant, even away from the FP point, since it is decoupled from the other fields, and we can seemingly conclude to the invariance of the reduced theory.

The reason why this argument is not totally correct is that, to perform  the reduction of 
${\cal L}'$ to the physical degrees of freedom, we must use the equations of motion of 
${\cal L}'$, rather than the equations of motion of ${\cal L}$. Thus, the correct derivation is the one that we have given above, in which we have have eliminated  from ${\cal L}'$ the non-dynamical variables, using their equations of motion derived from   Lagrangian 
${\cal L}'$ itself. In the end, however, the  only difference between using the equations of motion of ${\cal L}'$ or that of ${\cal L}$ is that with the former procedure (which in principle is the correct one)  both the term linear in $(A,C)$ and the terms quadratic in $(A,C)$ cancel exactly, as we checked explicitly. In contrast, with the (a priori incorrect) procedure of using the equations of motion obtained varying ${\cal L}$  to eliminate the non-dynamical variables from  ${\cal L}'$,  the terms linear in $A,C$ still cancel while those quadratic in $A,C$ do not. 
Since these quadratic term are  decoupled from the physical fields $\{\Phi,\eps_i,\hTTij\}$ they are  anyhow irrelevant, so in the end even the simpler procedure gives the correct answer.

\section{Hidden symmetry in the {\Stu} formalism}\label{sect:Stu}

An equivalent and instructive derivation of the existence of  this hidden symmetry can  be given by making use of the {\Stu} formalism.
We first introduce as usual a {\Stu} vector field ${\uppi}_{\mu}$,  performing  the replacement
\be\label{stuck1}
\hmn\ra \hmn  +\frac{1}{m_g}(\pam{\uppi}_{\nu}+\pan{\uppi}_{\mu})
\ee
in the action for linearized massive gravity. 
The linearized massive theory is then trivially invariant under the combined transformation
\bees
\hmn&\ra&\hmn  -(\pam\xin+\pan\xim)\, ,\\
{\uppi}_{\mu}&\ra& {\uppi}_{\mu}
+m_g\xim\, .\label{pitran}
\ees
We next introduce the {\Stu} scalar ${\uppi}$, replacing further
\be
{\uppi}_{\mu}\ra{\uppi}_{\mu}+\frac{1}{m_g}\pam{\uppi}\, .
\ee
The theory then also acquires  a $U(1)$ gauge symmetry,
\be\label{sympi}
\uppi_{\mu}\ra \uppi_{\mu}-\pam\theta\, ,\qquad
\uppi\ra\uppi +m_g\theta\, ,
\ee
while \eq{pitran} becomes
\be
{\uppi}_{\mu}+\frac{1}{m_g}\pam{\uppi}\ra {\uppi}_{\mu}+\frac{1}{m_g}\pam{\uppi}
+m_g\xim\, .\label{pitran2}
\ee
The overall  {\Stu} replacement is therefore
\be\label{Stutot}
\hmn\ra \hmn  +\frac{1}{m_g}(\pam{\uppi}_{\nu}+\pan{\uppi}_{\mu})
+\frac{2}{m_g^2}\pam\pan{\uppi}\, .
\ee
Writing $\xim$ as in \eq{gaugetran}, the transformation of the {\Stu} fields under linearized diffeomorphisms, \eq{pitran2}, becomes
\be\label{transfuppi}
{\uppi}_0\ra {\uppi}_0+m_g(A-\dot{C})\, ,\qquad
{\uppi}_i\ra {\uppi}_i +m_g B_i\, ,
\ee
\be
{\uppi}\ra{\uppi}+m_g^2 C\, ,
\ee
while in terms of the harmonic variables the {\Stu} transformation (\ref{Stutot}) reads
\be
\psi\ra \psi +\frac{1}{m_g}\dot{\uppi}_0+\frac{1}{m_g^2}\ddot{\uppi}\, ,\qquad
\phi\ra \phi-\frac{1}{3m_g^2}\n^2\uppi\, ,\label{Str1}
\ee
\be
\gamma\ra\gamma +\frac{1}{m_g}\uppi_0+\frac{2}{m_g^2}\dot{\uppi}\, ,\qquad
\lambda\ra\lambda +\frac{2}{m_g^2}\uppi\, ,\label{Str2}
\ee
\be
\beta_i\ra\beta_i+\frac{1}{m_g}\dot{\uppi}_i\, ,\qquad
\eps_i\ra\eps_i+\frac{2}{m_g}\uppi_i\, .\label{Str3}
\ee
We can now see how the hidden symmetry emerges in this formalism.  When we  perform the replacement  (\ref{Str1})-(\ref{Str3})
into the Lagrangian \eq{LmassiveG}  we obtain a new Lagrangian that depends both on the metric and on the {\Stu} fields. We then write down the  equations of motion of this Lagrangian with respect to $\gamma$, $\psi$ and $\n^2\lambda$ and use them to eliminate these variables from the theory.
Obviously, since the replacements (\ref{Str1}) and (\ref{Str2}) are patterned after the gauge transformation of the scalar fields, 
this is formally the same computation  already discussed
in sect.~\ref{sect:hidden}, with $\uppi$ playing the role of $-m_g^2C$ and $\uppi_0 $ of $-m_g(A-\dot{C})$. Thus  the result is  that, after elimination of the variables $\psi$, $\gamma$ and $\n^2\lambda$, the {\Stu} fields $\uppi$ and $\uppi_0$ disappear from the action, just as all terms involving $A$ and $C$ canceled in the  computation discussed
in sect.~\ref{sect:hidden}.
In conclusion, after reduction to the five physical degrees of freedom of the massive theory, the {\Stu} fields $\uppi_0$ and $\uppi$ drop from the action, and are no longer  needed for restoring the scalar part of the gauge symmetry. 
In contrast, $\uppi_i$ remains in the vector sector, where it is needed to obtain invariance under the transformation parametrized by $B_i$.

Observe also that, if we use the pair $(\bar{A},C)$ as independent variables, 
the {\Stu} fields $\uppi_0$ and $\uppi$ take care separately of the transformations parametrized by $\bar{A}$ and $C$,
\be
{\uppi}_0\ra {\uppi}_0+m_g\bar{A}\, ,\qquad
{\uppi}\ra{\uppi}+m_g^2 C\, .
\ee
In this  formulation, where the transformation parametrized by $C$ is simply
$\hmn\ra\hmn-2\pam\pan C$, we immediately recognize that
the hidden symmetry parametrized by $C$ is nothing but the  
$U(1)$ gauge symmetry (\ref{sympi}). Indeed, 
from \eq{stuck1}, the transformation $\uppi_{\mu}\ra \uppi_{\mu}-\pam\theta$ induces on the metric the transformation $\hmn\ra\hmn-2\pam\pan C$ with 
$C=\theta/m_g$.  The reason for performing a further {\Stu} transformation, introducing  the {\Stu} scalar $\uppi$, was to cancel this variation.
However we have seen that, after reduction to the physical degrees of freedom, the 
transformation $\hmn\ra\hmn-2\pam\pan C$  becomes a symmetry of the theory. Thus,  after performing the {\Stu} transformation (\ref {stuck1}) and eliminating the non-physical components of the metric, the $U(1)$ gauge symmetry is already there,
without the need of introducing   further the  {\Stu} scalar $\uppi$.

\section{(3+1) decomposition and Bardeen's variable in  electrodynamics}\label{sect:electro}

It is interesting to compare the above results with a similar analysis performed  both in massless and in  massive electrodynamics. We  begin  with the massless theory and apply the harmonic decomposition to the electromagnetic field,  writing the gauge field $A_{\mu}$ as
\be\label{decompAmu}
A_0=\psi\, ,\qquad A_i=v_i+\pa_i\lambda\, ,
\ee
where $v_i$ is a transverse vector, $\pa_i v^i=0$. Under a gauge transformation
$A_{\mu}\ra A_{\mu}-\pam\theta$ we have
\be
\psi\ra\psi-\dot{\theta}\, ,\qquad \lambda\ra\lambda -\theta\, ,
\ee
while $v_i$ is invariant (similarly to the fact that in linearized gravity $h_{ij}^{\rm TT}$ is invariant). We can then define a ``Bardeen variable"
\be
\Psi=\psi-\dot{\lambda}\, ,
\ee
which is gauge invariant. The four degrees of freedom of $A_{\mu}$ have therefore been rearranged into three gauge-invariant degrees of freedom $\{\Psi,v_i\}$, and one  pure-gauge degree of freedom, that can be taken to be $\lambda$ or $\psi$. 
Inverting \eq{decompAmu} we get
\be
\lambda= \n^{-2}(\pa_iA^i)\, ,\qquad v_i=A_i-\pa_i\n^{-2}(\pa_jA^j)\, .
\ee
Thus, as in linearized gravity, these harmonic variables are non-local function of the gauge field $\Am$ and the  non-locality disappears in the gauge  $\pa_iA^i=0$. 
We now couple the electromagnetic field to an external current $j^{\mu}$, and by analogy with \eqs{T00}{T0i} we write
\be
j_0=\rho\, ,\qquad
j_i=S_i+\pa_iS\, ,
\ee
with $\pa_i S^i=0$. Current conservation implies $\n^2S=\dot{\rho}$.
After some integrations by parts and use of this conservation equation, the Lagrangian density for the electromagnetic field coupled to an external current can be written as
\bees
{\cal L}&=&  -\frac{1}{4}\Fmn\FMN-j_{\mu}A^{\mu}\nn\\
&=&  \frac{1}{2}\[ \pa_i\Psi\pa^i\Psi -\pam v_i\paM v^i \]+
(\rho\Psi-S_i v^i )\, .
\ees
As expected, the Lagrangian density of the massless theory depends only on the gauge-invariant variables $\Psi$ and $v_i$. The variations with respect to $\Psi$ and $v_i$ give, respectively,
\be
\n^2\Psi =\rho\, ,\qquad
\Box v^i=S^i\, .
\ee
This shows that $v^i$ describes the two radiative degrees of freedom of the massless photon, while $\Psi$ is a physical but non-radiative degree of freedom.

We next consider the massive theory defined by the Proca Lagrangian density
\be
{\cal L}=  -\frac{1}{4}\Fmn\FMN-\frac{1}{2}m_{\g}^2A_{\mu}A^{\mu}\, .
\ee
Expressing it in harmonic variables we get ${\cal L}={\cal L}_{\rm scalar}
+{\cal L}_{\rm vector}$ with
\bees
{\cal L}_{\rm scalar}&=&  \frac{1}{2}\[ \pa_i\Psi\pa^i\Psi+
m_{\g}^2 (\psi^2-\pa_i\lambda\pa^i\lambda )\] \,,\\
{\cal L}_{\rm vector}&=&  -\frac{1}{2} \[ \pam v_i\paM v^i
+m_{\g}^2 v_iv^i
\] \,.
\ees
In the vector sector the variation with respect to $v_i$ gives the expected massive KG equation,
\be
(\Box-m_{\g}^2)v^i=0\, .
\ee
The Lagrangian in the scalar sector now depends separately on $\psi$ and $\lambda$, rather than only on the gauge-invariant combination $\Psi$. A  difference with respect to linearized gravity with the FP mass term is that now the variable $\psi$ enters quadratically.
We use  $\lambda$ and $\psi$ as independent variables. Clearly $\psi$ is non dynamical, since (recalling that $\Psi=\psi-\dot{\lambda}$) it enters the action without time derivatives. The variations with respect to $\psi$ and $\lambda$ give, respectively,
\be\label{qed1}
\n^2\psi -m_{\g}^2\psi =\n^2\dot{\lambda}\, ,
\ee
\be\label{qed2}
\ddot{\lambda}+m_{\g}^2\lambda=\dot{\psi}\, .
\ee
Taking the time derivative of \eq{qed1} and replacing $\dot{\psi}$ from \eq{qed2} we get
\be
(\Box-m_{\gamma}^2)\lambda =0\, .
\ee
This shows that the radiative degree of freedom in the scalar sector is $\lambda$, rather than the gauge-invariant variable $\Psi$ (an obvious result,  since we see
from \eq{decompAmu} that $\lambda$ corresponds to the longitudinal polarization). The variable $\Psi$ is instead determined by $\lambda$ through the equation
$(\n^2-m_{\gamma}^2)\Psi=m_{\gamma}^2\dot{\lambda}$, which is obtained writing $\psi=\Psi+\dot{\lambda}$ in \eq{qed1}.
Since $\lambda$ transforms as $\lambda\ra \lambda -\theta$, even after eliminating the non-dynamical variable $\psi$
no gauge symmetry appears in the Proca formulation of massive electrodynamics. The ``hidden" gauge symmetry is really a peculiarity of linearized massive gravity.

It is also instructive to see how the Lorentz invariance of massive electrodynamics is recovered using the harmonic  variables. 
Under an infinitesimal boost we have $A_0\ra A_0+\omega_{0i}A^i$ and $A^i\ra A^i+{\omega_0}^iA_0$. Then, proceeding as we did in sect.~\ref{sect:Lortransfor}, we find that under boosts
\bees
\d\psi&=&{\omega_{0}}^i\, (v_i+\pa_i\lambda)\, ,\\
\d\lambda&=&{\omega_{0}}^i\,\n^{-2}[\pa_i\Psi+\dot{v}_i]\, ,\\
\d v_i&=&-{\omega_{0}}^j \,\n^{-2}\, \[(\pa_i\pa_j-\d_{ij}\n^2)\Psi+\pa_i\dot{v}_j\]\, ,
\ees
and therefore (recalling that $\d\dot{\lambda}=\d(\pa_0\lambda)={\omega_{0}}^i\pa_i\lambda +\pa_0(\d\lambda)$)
\be
\d\Psi={\omega_{0}}^i\,  \n^{-2}\(\Box v_i-\pa_i\dot{\Psi}\)\, .
\ee
Observe that, just as in the gravitational case, the gauge-invariant variables $\Psi$ and $v_i$ transform among themselves under Lorentz transformations, but their transformation involve the non-local operator $\n^{-2}$.

Using  $\d(\pa_i\Psi)=\d(\pa_i)\Psi+\pa_i\d\Psi=\omega_{0i}\dot{\Psi}+
\pa_i\d\Psi$ we find
\be
\d\(\pa_i\Psi\pa^i\Psi\) = -2{\omega_{0}}^i\Psi\Box v_i\, ,
\ee
(plus a term $\pa_i\pa_0 [(1/2){\omega_{0}}^i\Psi^2]$ which is a total derivative). Similarly, neglecting total derivatives, we get
\be
\d\(\pam v_i\paM v^i\)=-2{\omega_{0}}^i\Psi\Box v_i\, ,
\ee
and
\be
\d (\psi^2-\pa_i\lambda\pa^i\lambda )=\d (v_iv^i)=
2{\omega_{0}}^i\Psi v_i\, .
\ee
Therefore, modulo total derivatives, 
\be
\d{\cal L}_{\rm scalar}=-\d{\cal L}_{\rm vector}=-\omega_{0i}\Psi
(\Box-m_g^2)v^i\, .
\ee
Thus the variation of ${\cal L}_{\rm scalar}$ cancels the variation of
${\cal L}_{\rm vector}$  and the total action  is Lorentz-invariant, as it should. Observe  that  $S_{\rm scalar}$ and $S_{\rm vector}$ are not separately Lorentz invariant
(unless we impose the equation of motion for $v^i$). 

\section{Radiative degrees of freedom and hidden gauge symmetry in De~Sitter space}\label{sect:DS}

In this section we consider massive gravity
linearized  over a  de~Sitter background.  We start from the Einstein-Hilbert action with a cosmological constant supplemented with a FP mass term, and we work for generality in $d$ spatial dimensions. Then
\be
S=\int d^{d+1}x\, \sqrt{-g}\, \[ \frac{2}{\kappa^2}(R-2\Lambda)-
\frac{m_g^2}{2}\(\hmn\hMN-h^2\)\]\, ,
\ee
where $\hmn$ is defined from
$\gmn=\gBmn+\kappa\hmn$. We write the
background metric in the form $\gBmn=(-1, a^2(t)\d_{ij})$, with $H=\dot{a}/a={\rm const}$.
The linearization of the action gives~\cite{Hinterbichler:2011tt}
\bees
S_2&=&\int d^{d+1}x\, \sqrt{-\bar{g}}\, \Bigl[
 -\frac{1}{2}\bar{D}_{\alpha}\hmn\bar{D}^{\alpha}\hMN  \nn \\ &&
 +\bar{D}_{\alpha}\hmn\bar{D}^{\nu}h^{\mu\alpha}
 -\bar{D}_{\mu}h\bar{D}_{\nu}\hMN
 +\frac{1}{2}\bar{D}_{\mu}h\bar{D}^{\mu}h\nn\\
 &&+\frac{\bar{R}}{d+1}\(\hmn\hMN-\frac{1}{2}h^2\)
 -\frac{m_g^2}{2}\(\hmn\hMN-h^2\) \Bigr]\, , \nn \\
\ees
where $\bar{D}_{\mu}$ is the covariant derivative with respect to the background metric $\gBmn$, and $\bar{R}$ is the Ricci scalar of the background. In de~Sitter space with $d$ spatial dimensions $\bar{R}=d(d+1)H^2$, and the Einstein equations for the background  fix it in terms of the cosmological constant, $\bar{R}=2[(d+1)/(d-1)]\Lambda$, so $H^2=2\Lambda/[d(d-1)]$.

As is well known,  if $m_g$ and $H$ satisfy the Higuchi bound \cite{Higuchi:1986py}
$m_g^2>(d-1)H^2$  
this theory has no ghost and five physical degrees of freedom: one in the scalar sector, two in the vector sector and two in the tensor sector. For 
$m_g^2<(d-1)H^2$ the degree of freedom in the scalar sector becomes a ghost, while at the special point $m_g^2=(d-1)H^2$ the scalar disappears and we  remain with just four propagating degrees of freedom. The theory at the special point  $m_g^2=(d-1)H^2$ is known as ``partially massless", and acquires an extra gauge symmetry of the 
form~\cite{Deser:2001pe,Deser:2001us,Deser:2001wx}
\be
\d\hmn=\bar{D}_{\mu}\bar{D}_{\nu}\alpha+\frac{m_g^2}{d-1}\a\gmn\, ,
\ee
parametrized by a function $\a(x)$.
It is interesting to recover these results using the $(3+1)$ decomposition, and 
see if the hidden symmetry found in the linearization over  Minkowski persists.
We limit ourselves to  the scalar sector of the theory. We  parametrize the metric fluctuations in the scalar 
sector as
\be\label{hijpsiDS}
h_{00}=2\psi\, ,\qquad\,
h_{0i}=\pa_i\gamma\, ,
\ee
\be\label{hijphilambdaepsDS}
h_{ij}= -2a^2\phi\d_{ij}+\(\pa_i\pa_j-\frac{1}{d}\d_{ij}\n^2\)\lambda\, .
\ee
We will use the sum over repeated spatial lower indices to denote contractions performed with the flat Minkowski metric, while contractions between upper and lower indices are done with the background FRW metric, so
$\pa_i\pa_i= \delta^{ij}\pa_i\pa_j$, 
$\pa_i\pa^i=a^{-2}\delta^{ij}\pa_i\pa_j$,
and
$\n^2\equiv \pa_i\pa_i$
still denotes the flat-space Laplacian. The factor $1/d$
in \eq{hijphilambdaepsDS} ensures that the operator acting on $\lambda$ is traceless in $d$ spatial dimensions.\footnote{The functions $E$ and $B$ commonly used in the literature are related to $\gamma$ and $\lambda$ by $\gamma=aB$ and $\lambda=2a^2E$.} The massless theory is invariant 
under linearized diffeomorphisms 
\be
\hmn\ra\hmn -(\bar{D}_{\mu}\xi_{\nu}+\bar{D}_{\nu}\xi_{\mu})\, .
\ee
In the scalar sector we write again $\xi_0=A$ and $\xi_i=\pa_iC$.
Then the gauge transformations of the scalar functions are
\bees
&\psi\ra \psi -\dot{A}\, ,\quad
&\phi\ra \phi+\frac{1}{d}\pa_i\pa^i C- HA\, ,\label{tr1dS}\\
&\lambda\ra\lambda -2C\, ,\quad
&\gamma\ra\gamma -A-\dot{C}+2HC\, .
\label{tr2dS}
\ees
In three-dimensional  Minkowski space these  transformations reduce to \eqs{tr1}{tr2}, as they should. The Bardeen variables in FRW read 
\bees
\Phi &=&-\phi-\frac{1}{2d}\pa_i\pa^i\lambda +H \gamma
-\frac{H}{2}(\dot{\lambda}-2H\lambda)\, ,\label{PhiDS}\\
\Psi&=&\psi-\dot{\gamma}+\frac{1}{2}(\ddot{\lambda}-2H\dot{\lambda})\, .
\label{PsiDS}
\ees
We find useful to rewrite them as
\begin{equation}  
\Phi = \Phi_0 + H \zeta \, , \hspace{1cm}  \Psi = \psi - \dot{\zeta} \, ,
\end{equation}
where 
\begin{equation}  
\Phi_0 \equiv - \phi - \frac{1}{2d} \pa_i\pa^i\lambda 
\end{equation}
is the Bardeen variable in Minkowski space and
\begin{equation}  
\zeta \equiv \gamma - \frac{1}{2} \dot{\lambda} + H\lambda \, .
\end{equation}
Observe that under  gauge transformations  $\zeta \to \zeta - A$. From \eqs{tr1dS}{tr2dS} it is straightforward to show that $\Phi$ and $\Psi$ are invariant under linearized diffeomorphisms. 
Using \eqst{hijpsiDS}{hijphilambdaepsDS} we find that the action in the scalar sector 
is\footnote{We have checked that for $m_g=0$  \eq{LmassiveGdeSit} agrees with the action given in eq.~(61) of
Rubakov and Tinyakov (RT)
\cite{Rubakov:2008nh}, after performing the change of notation $\psi=\varphi_{\rm RT}$,
$\gamma=aB$, $\lambda =2a^2E$ and 
$\phi +(1/2d)\pa_i\pa^i\lambda =\psi_{\rm RT}$ and going from cosmic to conformal time. Observe however that our expression given in terms of Bardeen's variables is much simpler.
The mass term examined in \cite{Rubakov:2008nh} is instead different from the one that we are considering, since even for a de~Sitter background it is constructed writing the FP mass term for the variable  $\gmn-\emn$, rather than for $\gmn-\gBmn$, and in this case the theory has a ghost, as shown in \cite{Rubakov:2008nh}.} 

\bees {\cal S}_{2,\rm scalar}=2(d-1)\int d^{d+1}x\, a^d
\Bigl[  -d(\dot{\Phi}+H\Psi)^2 
\nn\\ 
  +(d-2) \pa_i\Phi\pa^i\Phi -2\pa_i\Psi\pa^i\Phi\Bigr]\nn\\
\hspace*{-13mm}+\frac{m_g^2}{2} \int d^{d+1}x\, a^d \Bigl[ 
4d(d-1)\phi^2+8d\phi\psi
\nn\\  +2\pa_i\gamma\pa^i\gamma 
-\frac{d-1}{d} (\pa_i\pa^i\lambda)^2 \Bigr]\, .\label{LmassiveGdeSit}
\ees
Setting $a(t)=1$ and $d=3$, we recover the flat-space result (\ref {LmassiveG}). 
We now  eliminate the non-dynamical degrees of freedom from the action and  identify the variable describing the radiative degree of freedom.\footnote{In App.~\ref{app:compare} we compare 
our discussion with the results presented in  \cite{Alberte:2011ah}, where the equations of motion of the theory have been studied  using  Bardeen's variables, and a Klein-Gordon equation is obtained for the combination $a^{-2}(\Phi+\Psi)$. As we discuss in
App.~\ref{app:compare}, this however does not mean that $a^{-2}(\Phi+\Psi)$ is the radiative degree of freedom of the theory in the scalar sector.}
As independent variables we use
the set $\{\Phi_0, \psi, \gamma, \lambda \}$, which is the same that we used in the Minkowski space. 
Note that in de~Sitter $\Phi$  contains $\dot{\lambda}$ and therefore the change of variables from $\phi$ to $\Phi$ would not be legitimate, see footnote~\ref{foot:change}.

We now write the action (\ref{LmassiveGdeSit}) using these variables.  It is useful to rewrite 
the second and third term using the relation
\bees  \label{eq:trick1}
\int dt\, a^d\[ (d-2)\partial_i \Phi\partial^i\Phi - 2 \partial_i \Phi \partial^i \Psi \]= \nn\\ - 
\int dt\, a^d\frac{2}{H} \partial_i \Phi \partial^i \left( \dot{\Phi} + H\Psi \right) \, ,
\ees
which is proved writing $2a^d\pa_i\Phi\pa^i\dot{\Phi}=2a^{d-2}\pa_i\Phi
\pa_i\dot{\Phi}=a^{d-2}\pa_0(\pa_i\Phi\pa_i\Phi)$ and integrating by parts.
We also observe that
$\dot{\Phi} + H\Psi = \dot{\Phi}_0 + H\psi$. 
Then we get
\bees
&&S_{2,\rm scalar}  =  2(d-1)\int d^{d+1} x \, a^d  \nn\\
&& \times \left[ -d \left( \dot{\Phi}_0 + H \psi \right)^2 - \frac{2}{H} \partial_i \left( \Phi_0 + H\gamma \right) \partial^i \left( \dot{\Phi}_0 + H \psi \right)  \right] \nn \\
&& + 2(d-1)\int d^{d+1} x \, a^d \,(\pa_i\pa^i\lambda) \\
&& \times  \left[ \left( \ddot{\Phi}_0 + H \dot{\psi} \right) + d H \left( \dot{\Phi}_0 + H\psi \right) + m_g^2 \Phi_0 - \frac{m_g^2}{d-1} \psi  \right] \nn  \\
&&+ m_g^2\int d^{d+1} x \, a^d \left[ 2d(d-1)\Phi_0^2 - 4d\psi\Phi_0 + \partial_i \gamma \partial^i \gamma \right] \nn \, ,
\ees
where we have collected the terms proportional to $\pa_i\pa^i\lambda$ since it is a Lagrange multiplier. 
The equation of motion with respect to $\gamma$ is
\be
\gamma = \frac{2(d-1)}{m_g^2}\, \(\dot{\Phi}_0+H\psi\)\, .
\ee
So, just as in the flat-space case, $\gamma$ is a non-dynamical variable that can be eliminated using its own equation of motion, and we obtain
\bees 
\label{eq:preaction}
&&S_{2,\rm scalar} =  2(d-1)\int d^{d+1}x \, a^d \nn \\ && \times \Bigl[ -d \left( \dot{\Phi}_0 + H \psi \right)^2 - 2H^{-1} \partial_i \Phi_0 \partial^i \left( \dot{\Phi}_0 + H \psi \right) \nn\\
&&\hspace*{15mm}
- 2\frac{(d-1)}{m_g^2}  \partial_i \left( \dot{\Phi}_0 + H\psi \right)  
\partial^i \left( \dot{\Phi}_0 + H\psi \right)  \Bigr]\nn\\   
& & + 2(d-1)\int d^{d+1}x \, a^d \, (\pa_i\pa^i\lambda) \Bigl[ \left( \ddot{\Phi}_0 + H \dot{\psi} \right) \nn\\
&&\hspace*{10mm}+ d H \left( \dot{\Phi}_0 + H\psi \right) + m_g^2 \Phi_0 - \frac{m_g^2}{d-1} \psi \Bigr] 
\nn\\
& & + m_g^2\int d^{d+1}x \, a^d \left[ 2d(d-1)\Phi_0^2 - 4d\psi\Phi_0 \right]  \, .
\ees
We next rescale our variables 
\begin{equation}  
\{ \psi, \Phi_0, \lambda \} \to a^{-(d-1)}\{ \psi, \Phi_0, \lambda \} \, ,
\end{equation}
define an effective mass
\begin{equation}  
M^2 \equiv m_g^2 - (d-1)H^2 \, ,
\end{equation} 
and  use 
\be
\chi \equiv \psi - (d-1)\Phi_0\, 
\ee
as an independent variable instead of $\psi$. We get
\bees 
&&S_{2,\rm scalar} =  2(d-1)\int d^{d+1}x \, a^{-d+2} \nn \\ &&\times \Bigl[ -d \left( \dot{\Phi}_0 + H \chi \right)^2  - 2H^{-1} \partial_i \Phi_0 \partial^i \left( \dot{\Phi}_0 + H \chi \right) 
 \nn\\
&&\hspace*{15mm} - \frac{2(d-1)}{m_g^2}  \partial_i \left( \dot{\Phi}_0 + H \chi \right) 
\partial^i \left( \dot{\Phi}_0 + H \chi \right) \Bigr] \nn\\
& & + 2(d-1)\int d^{d+1}x \, a^{-d+2}   (\pa_i\pa^i\lambda)\nn \\ &&\hspace*{25mm}\times \left[ \ddot{\Phi}_0 + H \dot{\Phi}_0 + H\dot{\chi} - \frac{M^2}{d-1} \chi \right]\nn\\
&&+ 2(d-1)\int d^{d+1}x \, a^{-d+2} \[  - dm_g^2 \Phi_0^2 - \frac{2d}{d-1}m_g^2\chi\Phi_0  \] \, . \nn \\
\label{actPhi0chi}
\ees
For $M \neq 0$ we now trade in the action $\Phi_0$ for the new variable $\Omega$ defined by
\bees \label{eq:Omegadef}
&&\Omega(t,\vx) \equiv \Phi_0(t,\vx) + H \left( \int_{t_0}^t {\rm d}t' \chi(t',\vx) \right)\nn\\&&
+ \frac{(d-1)H}{M^2} \left[ \dot{\Phi}_0(t_0,\vx) + H\Phi_0(t_0,\vx) + H\chi(t_0,\vx) \right] \, , \nn\\
\ees
where $t_0$ is the time at which the initial conditions are given.  The first two terms are chosen so that 
\be\label{OdotPhichi}
\dot{\Omega}=\dot{\Phi}_0 + H \chi\, .
\ee
The last term in brackets is time-independent and is a functional of the initial data  $\Phi_0(t_0,\vx)$, $ \dot{\Phi}_0(t_0,\vx)$ and $\chi (t_0,\vx) $. The reason for this specific choice of the time-independent term will become clear below.
Note that the  initial data $\{ \Phi_0(t_0), \dot{\Phi}_0(t_0)\}$ and $\{\chi(t_0), \dot{\chi}(t_0) \}$ fully determine $\{ \Omega(t_0), \dot{\Omega}(t_0) \}$,
\bees 
\Omega(t_0) &=& \frac{m_g^2}{M^2}\Phi_0(t_0) + \frac{(d-1)H}{M^2} \left( \dot{\Phi}_0(t_0) +  H\chi(t_0) \right)  \, ,\nn\\ \dot{\Omega}(t_0) &=& \dot{\Phi}_0(t_0) + H\chi(t_0)  \, 
\ees
(where, for notational simplicity, here and in the following we do not write explicitly the $\vx$ dependence). Inverting we get
\bees  \label{eq:icrel}
\Phi_0(t_0) &=& \frac{M^2}{m_g^2}\Omega(t_0) - \frac{(d-1)H}{m_g^2} \dot{\Omega}(t_0)  \, , \nn\\ \dot{\Phi}_0(t_0) &=& \dot{\Omega}(t_0) - H\chi(t_0)  \, .
\ees
The occurrences of
$\Phi_0$ other than in the combination (\ref{OdotPhichi})
bring in terms  that are non-local in time. Using \eq{eq:icrel} we get in fact
\bees \label{eq:PhiofOm}
\Phi_0(t) &=& \Omega(t) - H \left( \int_{t_0}^t {\rm d}t' \chi(t') \right) \nn\\ &&- \frac{(d-1)H}{m_g^2} \left( \dot{\Omega}(t_0) + H\Omega(t_0) \right) \, .
\ees
However, in these new variables the constraint imposed by the Lagrange multiplier $\pa_i\pa^i\lambda$ reads
\begin{equation}  
\chi = \frac{d-1}{m_g^2} \left(  \ddot{\Omega} + H \dot{\Omega} \right) \, ,
\end{equation} 
so integrating out $\lambda$ and eliminating $\chi$ from the action makes
\eq{eq:PhiofOm}  local in time, 
\begin{equation}  
\Phi_0 = \frac{M^2}{m_g^2}\Omega - \frac{(d-1)H}{m_g^2} \dot{\Omega} \, .
\end{equation}
The time-independent term in the definition of $\Omega$, \eq{eq:Omegadef}, was precisely chosen so that 
time-independent
part in this expression vanishes. The action now depends only on $\Omega$ and, after some integrations by parts and rearrangements we find
\bees
&&S_{2,\rm scalar} = \\ 
&& -2d(d-1)\frac{M^2}{m_g^2} \int d^{d+1}x \, a^{-d+2} \left[ \partial_{\mu}\Omega \partial^{\mu}\Omega + M^2 \Omega^2 \right] \, . \nn 
\ees
We  now rescale back
\begin{equation}  
\Omega \to a^{d-1}\Omega \, ,
\end{equation}
so that in the action  we reconstruct  the volume form $\sqrt{-g}=a^d$,
\begin{equation}  \label{Omegaact}
S_{2,\rm scalar} = -2d(d-1)\frac{M^2}{m_g^2}\int d^{d+1}x \, a^d \left[ \partial_{\mu}\Omega \partial^{\mu}\Omega + m_g^2 \Omega^2 \right] \, . 
\end{equation}
This also has the effect of replacing the effective mass $M^2$ by $m_g^2$ inside the square brackets. 
In terms of the original variables, the dynamical mode is
\begin{eqnarray}  
&&\Omega(t)  =  \Phi_0(t)\nn\\&& + a^{-(d-1)}(t) H \int_{t_0}^t {\rm d}t' a^{d-1}(t') \left[ \psi(t') - (d-1) \Phi_0(t') \right] \nn\\
 & & + \frac{(d-1)H}{M^2} \(\frac{a(t)}{a(t_0)}\)^{-(d-1)} \nn \\ &&  \hspace*{15mm} \times\left( \dot{\Phi}_0(t_0) + H\Phi_0(t_0) + H\psi(t_0) \right)  \, . \label{defOme} 
\end{eqnarray} 
After an integration by parts, we can rewrite it as
\begin{eqnarray}  
&&\Omega(t) =  a^{-(d-1)}(t)\int_{t_0}^t {\rm d}t' a^{d-1}(t') \left( \dot{\Phi}_0(t') + H \psi(t') \right) 
\nn \\
&& + \(\frac{a(t)}{a(t_0)}\)^{-(d-1)}  \\ 
&& \times \left[ \frac{(d-1)H}{M^2} \left( \dot{\Phi}_0(t_0) + H\psi(t_0) \right) + \frac{m_g^2}{M^2} \Phi_0(t_0) \right]  \, . \nn\label{Omegaint}
\end{eqnarray}
\Eqs{Omegaact}{Omegaint} are the main result of this section, and  nicely display in a compact manner a number of known features of massive gravity in de~Sitter space. 
First of all we see that for $M^2>0$, i.e. for 
$m_g^2 > (d-1)H^2$, the kinetic term of the scalar field has the ``good" non-ghostlike sign and the mass term is non-tachyonic. For $M^2<0$ we  have a ghost instead (and the mass term in \eq{Omegaact} becomes tachyonic). For $M=0$ the action vanishes and the radiative degree of freedom in the scalar sector disappears.\footnote{Actually, the manipulations leading to \eq{Omegaact} were performed assuming $M\neq 0$. However, for $M=0$,  one can already eliminate $\chi$ using the constraint imposed by the Lagrange multiplier $\lambda$ in 
\eq{actPhi0chi}. This gives $H\chi =-(\dot{\Phi}_0+H\Phi_0)$. Replacing this expression for $\chi$ into the action gives $S_{2,\rm scalar}=0$, so in \eq{Omegaact} the limit $M^2\ra 0$ is smooth.} 
Observe also that we smoothly retrieve the flat space-time result in the $H \to 0$ limit. The action
(\ref{Omegaact})  also agrees with that found in \cite{deRham:2012kf} using a different route, namely embedding $D$-dimensional de~Sitter space in $(D+1)$-dimensional Minkowski space and using the {\Stu} formalism to isolate the helicity-0 mode.

To make contact with the well-known result given by
Deser and Waldron~\cite{Deser:2001wx},
for $M^2>0$ we can introduce a new variable 
\be
q_0(t)=2\[ d(d-1)\frac{M^2}{m_g^2}\]^{1/2} a^{d/2}\Omega(t)
\ee
and  the action (\ref{Omegaact}) can be rewritten as
\begin{equation}  \label{OmegaactDW}
S_{2,\rm scalar} = -\frac{1}{2}\int d^{d+1}x \,  \left[ \partial_{\mu}q_0 \partial^{\mu}q_0 + \(m_g^2-\frac{d^2H^2}{4}\) q_0^2 \right] \, , 
\end{equation}
which, specialized to $d=3$, reproduces the result of \cite{Deser:2001wx}. Note that using the variable $q_0$ the volume form $\sqrt{-g}=a^d$ has  been eliminated in favor of an additional contribution to the mass term.

Finally, having found the expression for the field that describes the radiative degree of freedom in the scalar sector, we can ask whether the gauge symmetry in the scalar sector is still preserved, as in Minkowski space. 
\Eq{Omegaint} shows explicitly that the dynamical variable $\Omega$  is invariant under gauge transformations parametrized by  $C$  since, under this transformation,  $\Phi_0(t)$ and $\psi(t)$ are invariant. In contrast,  under the transformation parametrized by $A$ we have
\begin{equation}  
\Phi_0 \ra\Phi_0+ HA \, , \hspace{1cm}   \psi\ra\psi -\dot{A}\, .
\end{equation}
Therefore, the terms in $\Omega(t)$ containing the combination $\dot{\Phi}_0(t) + H \psi(t)$ are invariant, while the term $\Phi_0(t_0)$  has a variation determined by $A(t_0)$, so
\be
\Omega(t,\vx)\ra\Omega(t,\vx)+\(\frac{a(t)}{a(t_0)}\)^{-(d-1)} \frac{m_g^2H}{M^2}A(t_0)\, .
\ee
We see that the function $\Omega(t,\vx)$ has a variation that depends only on the value of the function $A(t)$ at the initial time $t_0$, rather than on the whole function $A(t)$.
We therefore still get an invariance if we restrict to gauge functions $A(t)$ that vanish at the initial time chosen for assigning the initial conditions. Until now we have worked with $t_0$ arbitrary. A natural choice, however, is to chose $t_0=-\infty$, and to restrict to gauge functions $A(t)$ that vanish as $t\ra -\infty$. With this restriction on the initial value of the function $A(t)$, the variable $\Omega(t)$ is invariant under the gauge transformations parametrized by both $C$ and by $A$, and  therefore  the ``hidden symmetry"  that we found in massive gravity linearized over Minkowski space persists in massive gravity linearized over  de~Sitter space.

\section{Conclusions and summary}\label{sect:concl}

We conclude by  summarizing the main results and equations of this rather long paper.
The use of the $(3+1)$ decomposition of the metric and of Bardeen's variables provides a valuable tool for understanding various aspects of massive gravity. This formalism allows to identify and eliminate the non-dynamical degrees of freedom from the action,
working directly at the Lagrangian rather than Hamiltonian level,  and to write down a reduced action for  the radiative degrees of freedom.
We have shown how to carry out  this elimination procedure both in Minkowski and in de~Sitter space. For massive gravity linearized over Minkowski space (in $d$ spatial dimensions) we found that, in the scalar sector, the variable that describes the radiative degree of freedom of the massive graviton is the flat-space Bardeen variable 
\be
\Phi_0=-\phi-\frac{1}{2d}\n^2\lambda\, ,
\ee
with $\phi,\lambda$ defined by the harmonic decomposition of the metric,
\eq{hijphilambdaeps}.
 (In sects.~\ref{sect:harmo}-\ref{sect:hidden} we denoted this variable simply as $\Phi$. Here we reserve the notation $\Phi$ for the curved-space Bardeen potential). In terms of $\Phi_0$, for a mass term that has the Fierz-Pauli form and in generic $d$ spatial dimensions, the reduced theory  in the scalar sector is simply described by  a Klein-Gordon action,
\be\label{PhicovariantConcl}
{\cal S}_{2,\rm scalar} 
=-2d(d-1)\int d^{d+1}x\, (\pam\Phi_0\paM\Phi_0 +m_g^2\Phi_0^2)\, ,
\ee
and therefore $\Phi_0$ satisfies a KG equation,
\be\label{KGPhiSource3Concl1}
(\Box-m_g^2)\Phi_0=0\, ,
\ee
a result that was already found using the ADM formalism in \cite{Deser:1966zzb}
(see also \cite{Alberte:2010it}).
In the presence of external matter, we  found that this KG equation is sourced by the combination $(\rho- \n^2\sigma)$, see \eq{KGPhiSource3}.
This result is interesting and somewhat unexpected for various reasons. First of all, in the massless case the Bardeen variables describe a physical but non-radiative degree of freedom, as we recalled in sect.~\ref{sect:radiative}. It is surprising to see that, when we switch on a mass term, it is precisely the combination $\Phi_0$ that describes the radiative degree of freedom.
The real surprise, however, comes from the fact that a variable such as $\Phi_0$ is gauge-invariant under linearized diffeomorphisms. Thus,
despite the fact that the mass term breaks the gauge invariance of the theory, 
after elimination of the non-dynamical variables the scalar sector of massive gravity is still 
gauge invariant, in the sense discussed in sect.~\ref{sect:hidden}.  
Writing the gauge transformation in the form
$\hmn\ra\hmn  -(\pam\xin+\pan\xim)$, with
$\xi_0=A$, $ \xi_i=B_i+\pa_iC$,
the reduced theory is invariant under the gauge transformations parametrized by the scalar functions $A$ and $C$.  The crucial point for the existence of this symmetry is the fact that, after eliminating the non-dynamical degrees of freedom, the scalar sector can be written uniquely in terms of $\Phi_0$ (possibly including higher-derivative terms, if we are not at the FP point). This was not obvious a priori: in principle, one could have remained with a different non-gauge invariant field. This is indeed what happens in the vector sector, where  the propagating field is $\eps^i$, rather than the gauge-invariant combination $\Xi^i$. Thus, in the vector sector, the symmetry parametrized by $B_i$ is broken. 
Similarly, in massive electrodynamics in the scalar sector survives the longitudinal mode, which is not gauge-invariant. Thus, the existence of a hidden symmetry is a peculiar and  non-trivial property of the scalar sector of linearized massive gravity.

It is also interesting to explore the structure of the theory in these variables when the mass term deviates from the FP form. In this case the scalar sector can be described by two fields $(\Phi_0,\Gamma)$ and is governed by the Lagrangian density (\ref{PhiGammafull}),
which is second-order in the time derivatives. When the coefficients $b_1,b_2$ that parametrize the mass term satisfy $(b_1+b_2)/b_1b_2\leq 0$,
the combination $\Gamma$ given in terms of the metric in \eq{defGamma} is the ghost, while $\Phi_0$ is the healthy mode. Otherwise the ghost is obtained from a mixture of the Fourier modes of $\Phi_0$ and $\Gamma$.
Alternatively we can integrate out even $\Gamma$, at the price of a higher-derivative action. It is remarkable to see that in this case the complicated Lagrangian (\ref{PhiGammafull}) collapses to a simple covariant form for $\Phi_0$   
\be
{\cal L}_{2,\rm scalar} 
=
\a_1 \pam\Phi_0\paM\Phi_0 +
\a_2\,m_g^2\Phi_0^2 +\frac{\a_3}{m_g^2}\, 
(\Box\Phi_0)^2\, ,
\ee
with coefficients $\a_1,\a_2,\a_3$ given (for $d=3$) in
\eqs{a1a2b}{a3b}, and in particular $\a_3=0$ at the FP point.
The corresponding source term is given in \eq{miracle2}.
The appearance of such explicitly Lorentz-covariant structures 
is required by the fact that, in the end, the variables that describe the helicity-0 modes must have a Lorentz-invariant dispersion relation. Observe however that
$\Phi_0$ is {\em not} a scalar field  under Lorentz transformations, but is scalar only under spatial rotations, and the Lorentz invariance of the theory only emerges combining the scalar, vector and tensor sectors.

Finally, we have explored massive gravity linearized over de~Sitter using this formalism. In this case the elimination of the non-dynamical variables and the identification of the radiative degree of freedom in the scalar sector proved to be more subtle. The radiative degree of freedom is not given by the curved-space Bardeen variable $\Phi$ defined in \eq{PhiDS} (nor by the combination $(\Phi+\Psi)$, despite the fact that it satisfies a KG equation, see app.~\ref{app:compare}). Instead, it is described by the field $\Omega$ given in \eq{Omegaint}, which has the form
\be
a^{d-1}(t)\, \Omega(t,\vx)  =  \int_{t_0}^t {\rm d}t' a^{d-1}(t') (\dot{\Phi}_0 + H \psi ) (t',\vx) + f(t_0,\vx)\, ,
\ee
where $\Phi_0$ is still the flat-space Bardeen variable, $2\psi=h_{00}$, and $f(t_0,\vx)$ is a time-independent function which is fixed by the initial conditions imposed at the initial time $t_0$.
This expression is non-local in time. However, in the limit $H\ra 0$, $\Omega$ becomes local and smoothly reduces to $\Phi_0$ . In term of this variable, and for a FP mass term,  the reduced theory in the scalar sector is again described by a simple KG action,
\bees \label{OmegaactConc}
 \hspace*{-10mm}S_{2,\rm scalar} &=& -2d(d-1)\frac{[m_g^2 - (d-1)H^2]}{m_g^2} \nn\\
&& \hspace*{-5mm}\times
\int d^{d+1}x \, \sqrt{-g}\,  \left[ \partial_{\mu}\Omega \partial^{\mu}\Omega + m_g^2 \Omega^2 \right] \, . 
\ees
This expression nicely summarizes a number of known facts about massive gravity in de~Sitter:  (1) if $m_g^2 > (d-1)H^2$ (the Higuchi bound) we have a single healthy degree of freedom in the scalar sector with the correct sign for the kinetic term and a non-tachyonic mass term (we always assume $m_g^2>0$). (2)
for  $m_g^2 < (d-1)H^2$, the scalar degree of freedom becomes a ghost, and  (3)  at the special point $m_g^2 = (d-1)H^2$, corresponding to the so-called partially massless case, the scalar degree of freedom disappears from the spectrum of the theory. 

Furthermore, the variable $\Omega$ is invariant under scalar gauge transformations parametrized by $A$ and $C$ (with the extra condition that we restrict to gauge functions $A(t)$ that
vanish at the initial time $t_0$, which can be chosen to be $t_0=-\infty$), and therefore the ``hidden symmetry" that we have found in Minkowski space persists in de~Sitter.

Our analysis has always been performed without including the non-linearities of the theory. 
It would be interesting to generalize it  to the non-linear ghost-free  de~Rham-Gabadadze-Tolley model 
of massive gravity \cite{deRham:2010ik,deRham:2010kj}. Work on this is currently in progress.

\vspace*{5mm}

\noindent {\bf Acknowledgments}.
We thank Stanley Deser, Ruth Durrer and Antonio Riotto for useful comments on the manuscript.
Our work is supported by the Fonds National Suisse.

\appendix

\section{Diagonalization of the kinetic term in the scalar sector}\label{app:A}

In this appendix we study the diagonalization of the kinetic term of the action
(\ref{PhiGammafull}). 
To  study the existence of ghosts we consider at   first only the terms  that depend on time derivatives, i.e. we consider
the Lagrangian
\be
L=\int d^3x\, \[12\dot{\Phi}^2+\dot{\Gamma}\dot{\Phi}-
\frac{16(b_1+b_2)}{b_1b_2m_g^2}\pa_i\dot{\Phi}\pa^i\dot{\Phi}\]\, 
\ee
(so that the corresponding action is $S=\int dt\, L$).
It is convenient to perform a spatial Fourier transform of $\Phi(t,\vx)$ and
$\Gamma(t,\vx)$ and work with the modes
$\Phi_{\bf k}(t)$ and $\Gamma_{\bf k}(t)$, so
\be\label{LagL}
L=\int\, \frac{d^3k}{(2\pi)^3}\, 
\[ \frac{1}{2}\alpha_k\dot{\Phi}_{\bf k}^*\dot{\Phi}_{\bf k}+
\dot{\Gamma}_{\bf k}^*\dot{\Phi}_{\bf k}\]\, ,
\ee
where 
\be\label{AkBk}
\alpha_k= 24-\frac{32(b_1+b_2)|\vk|^2}{b_1b_2m_g^2}\,.
\ee
Since $\Phi(t,\vx)$ and
$\Gamma(t,\vx)$ are real, the momentum modes satisfy
$\Phi_{\bf k}^*=\Phi_{-\bf k}$ and
$\Gamma_{\bf k}^*=\Gamma_{-\bf k}$.
The corresponding conjugate momenta are
$\Pi_{\Phi_{\bf k}}=\d L/\d\dot{\Phi}_{\bf k}=\alpha_k\dot{\Phi}_{\bf k}^*
+\dot{\Gamma}_{\bf k}^*$ and 
$\Pi_{\Gamma_{\bf k}}=\d L/\d\dot{\Gamma}_{\bf k}=\dot{\Phi}_{\bf k}^*$, and the Hamiltonian is
\bees
\hspace*{2mm}H&=&\int\, \frac{d^3k}{(2\pi)^3}\, \[\Pi_{\Phi_{\bf k}}\dot{\Phi}_{\bf k}+
\Pi_{\Gamma_{\bf k}}\dot{\Gamma}_{\bf k}\] -L \nn\\
&=&\int\, \frac{d^3k}{(2\pi)^3}\, \[
\frac{1}{2}\alpha_k\dot{\Phi}_{\bf k}^*\dot{\Phi}_{\bf k}+
\dot{\Gamma}_{\bf k}^*\dot{\Phi}_{\bf k}
\]\, ,
\ees
so $H=L$. This Hamiltonian can be diagonalized writing it as
\be
H=\int\, \frac{d^3k}{(2\pi)^3}\,  \, \left[ \frac{1}{2\a_k}
\left| \a_k\dot{\Phi}_{\bf k}+\dot{\Gamma}_{\bf k}\right|^2
-\frac{1}{2\alpha_k}\left|\dot{\Gamma}_{\bf k}\right|^2\,\,\right] \, .
\ee
This shows that, independently of the sign of $\alpha_k$, the Lagrangian (\ref{LagL})
always has a ghost \cite{Rubakov:2008nh}. In particular,
if $\alpha_k>0$ the ghost is $\Gamma_{\bf k}$ while for 
$\alpha_k<0$ are  ghost-like the Fourier modes  
$\Gamma_{\bf k}+\alpha_k\Phi_{\bf k}$. We see from \eq{AkBk} that, if 
\be
\frac{b_1+b_2}{b_1b_2}\leq 0\, ,
\ee 
$\alpha_k$ is  positive for all $k$, so  the ghost is $\Gamma (t,\vx)$. For
$(b_1+b_2)/b_1b_2>0$ the situation is instead quite peculiar: there is a critical value 
$k_*\equiv |\vk_*|$ defined by $\alpha_{k_*}=0$, i.e.
\be
\( \frac{k_*}{m_g}\)^2=\frac{3}{4}\, \frac{b_1b_2}{b_1+b_2}\, , 
\ee
such that are ghost-like the  modes 
$\Gamma_{\bf k}$ with $|\vk|<k_*$ and the modes
$\a_k\Phi_{\bf k}+\Gamma_{\bf k}$ with $|\vk|>k_*$.
Thus, in  the two-parameter space  $(b_1,b_2)$ that parametrizes the mass term, 
there is a region 
where are ghost-like the Fourier  modes of a combination of metric components with momentum $k$ smaller than a critical value $k_*$, and the Fourier modes of a different combination of metric components with $k>k_*$, a situation which is quite unusual.

For  comparison with the masses of the vector and tensor modes, it is interesting to compute the masses of the two scalar modes  for $b_1$ and $b_2$ generic.  This can be conveniently done specializing to field configurations independent of $\vx$, i.e. to the Fourier modes with $\vk =0$. 
We also introduce canonically normalized fields $\Phi_N=(24)^{1/2}\Phi$ and
$\Gamma_N=(24)^{-1/2}\Gamma$. Then
the Lagrangian (\ref{PhiGammafull})  becomes
\bees
&& {\cal L}_{2,\rm scalar} [\Phi_N(t),\Gamma_N(t)] = \nn\\
&&\frac{1}{2}\dot{\Phi}_N^2+\dot{\Gamma}_N\dot{\Phi}_N
-\frac{1}{2}m_G^2 (\Phi_N^2+2\Gamma_N\Phi_N+\alpha \Gamma_N^2)\, ,\nn \\
\ees
where as usual $m_G^2\equiv b_1m_g^2$,
and 
\be
\alpha \equiv \frac{ 3(b_1+2b_2)}{2(b_1+b_2)}\, . 
\ee
Going to  Fourier space, we get
\be
S_{2,\rm scalar}=\frac{1}{2}\int \frac{d\omega}{2\pi}\, 
(\tilde{\Phi}_{-\omega},\tilde{\Gamma}_{-\omega}) M(\omega)
\(\begin{array}{c}
\tilde{\Phi}_{\omega}\\\tilde{\Gamma}_{\omega}
\end{array}\)\, ,
\ee
with
\be
M(\omega)= \(\begin{array}{cc}
\omega^2-m_G^2 &\omega^2-m_G^2 \\
\omega^2-m_G^2 & -\alpha m_G^2
\end{array}\)
\ee
Inverting this   quadratic form we find that the propagator has poles at $\omega^2=m_G^2$ and at 
\be\label{massG1A}
-\omega^2=\frac{ (b_1+4b_2)}{2 (b_1+b_2)} m_G^2\, ,
\ee
which therefore are the masses of the two degrees of freedom in the scalar sector. The negative sign in front of $\omega^2$ in \eq{massG1A}
reflects the fact that this mode is ghost-like, hence its kinetic term has the opposite sign compared to a standard kinetic term.  

\section{Derivation of $j^0=0, \pa_ij^i=0$ from the non-dynamical equations of motion}
\label{app:jnu}

As we mentioned in Sect.~\ref{sect:hidden},  defining $j^{\nu}$ as in \eq{defjnu},  the use of the non-dynamical equations of motion ensures that $j^0=0$ and $\pa_ij^i=0$.
To prove this result we write $j^{\nu}$ using the harmonic decomposition of the metric. For the temporal component  we get
\be
j^0=-(\n^2\gamma+6\dot{\phi})\, .
\ee
Using \eqs{elimgammaFP}{varpsiFP} we find $j^0=0$. Therefore, the condition
(\ref{condic1c2}) with $\nu=0$ is indeed a consequence of the non-dynamical equations
of motion.

For the spatial component $j^i$ the situation is different. In harmonic variables,
\bees
j^i&=&
b_1\(-\pa^i\dot{\gamma}-2\pa^i\phi+\frac{2}{3}\pa^i\n^2\lambda
+\frac{1}{2}\n^2\eps^i-\dot{\beta}^i\)
\nn\\ &&-2b_2(\pa^i\psi+3\pa^i\phi)\, .
\ees
This quantity depends both on the scalar functions $\psi,\phi,\gamma, \lambda$, and on the vectors $\eps^i$ and $\beta^i$. The part depending on the scalar functions vanishes upon use of the non-dynamical equations of motion (\ref{elimgammaFP}) and $\Gamma=0$. We remain with 
\be\label{ji}
j^i=\frac{1}{2} \(\n^2\eps^i-2\dot{\beta}^i\)\, . 
\ee
We found in \eq{conditionepsbeta} that indeed the equations of motion in the vector sector imply the condition $\n^2\eps^i-2\dot{\beta}^i=0$. However, to get this result we used the dynamical equation of motion (\ref{eqmotepsi}) obtained performing the variation with respect to $\eps^i$.
Thus, the use of the non-dynamical equations of motion  is not sufficient to derive the condition $j^{\mu}=0$. For this, it is necessary to use the full equations of motion, including a dynamical component. However, $\eps^i$ and $\beta^i$ are transverse vectors, so
from \eq{ji} it follows $\pa_i j^i=0$. This means that, at the FP point,
the non-dynamical equations are  sufficient to derive the conditions
\be
j^0=0\, ,\qquad \pa_ij^i=0\, , 
\ee
and therefore also 
$\pam j^{\mu}=0$.

\section{Equations of motion in terms of  $\Phi$ and $\Psi$ in de~Sitter}\label{app:compare}

In this appendix we compare our analysis with the discussion of the equations of motion in de~Sitter performed in  \cite{Alberte:2011ah}.
Taking the variation of the action (\ref{LmassiveGdeSit})
with respect to $\psi,\gamma,\lambda$ and $\phi$ 
one finds
\bees
\d\psi:&&\hspace*{-3mm}(d-1)\frac{1}{a^2}\n^2\Phi-d(d-1)H\(\dot{\Phi}+H\Psi\)=-dm_g^2\phi\, , \nn\\\label{dpsi}\\
\d\gamma:&&\dot{\Phi}+H\Psi=\frac{m_g^2}{2(d-1)}\,\gamma\, ,\label{dgamma2}\\
\d\lambda:&&(d-2)\Phi-\Psi=\frac{m_g^2}{2}\, \lambda\, ,\label{dlambda}\\
\d\phi:&&\ddot{\Phi}+dH\dot{\Phi}+H\dot{\Psi}+dH^2\Psi
+\frac{1}{da^2}\n^2\[\Psi -(d-2)\Phi\] \nn\\ &&=\frac{m_g^2}{d-1}\, 
\[ (d-1)\phi+\psi\].
\label{dphi}
\ees
Using
\eqs{dgamma2}{dlambda} one can eliminate  $\gamma$ and $\lambda$, and then the definitions of $\Phi$ and $\Psi$,
\eqs{PhiDS}{PsiDS}, allow us to express $\phi,\psi$ in terms of $\Phi$ and $\Psi$. Plugging the resulting expressions into \eqs{dpsi}{dphi}  and combining the equations one finds the coupled set of equations
\bees\label{AdPsi}
\frac{1}{a^2}\n^2(\Phi+\Psi) +dH(\dot{\Phi}+\dot{\Psi})+
d(d-3)H^2\Psi \nn\\ +\[2d(d-2)H^2-dm_g^2\]\Phi=0\, ,
\ees
and
\be\label{AdPhi}
\(\Box-m_g^2\) \[ a^{-2}(\Phi+\Psi)\]=0\, ,
\ee
where $\Box =(1/\sqrt{-g})\pam(\sqrt{-g}\, \gMN\pan)$ is the d'Alembertian in curved space on scalar functions.\footnote{In the derivation one uses the fact  that
in de~Sitter, for any scalar function $f$, we have
$-a^2\Box ( a^{-2}f) =\ddot{f}+(d-4)H\dot{f}-2(d-2)H^2f-a^{-2}\n^2f$.}
When  specialized to $d=3$,
\eqs{AdPsi}{AdPhi} agree with the result presented in \cite{Alberte:2011ah}, after the appropriate change of notation.\footnote{In particular, in \cite{Alberte:2011ah} our $\Phi$ is called $\Psi$  and our $\Psi$ is called $\Phi$, the metric signature is the opposite and the equations are written in conformal time. See also \cite{Comelli:2012db} for similar perturbation equations in the context of  bigravity.}
In flat space \eqs{AdPsi}{AdPhi} reduce to
\bees
&&\n^2(\Phi+\Psi)= dm_g^2\Phi\, ,\label{n2PhiPsi}\\
&&\(\Box-m_g^2\) (\Phi+\Psi)=0\, .\label{BoxPhiPsi}
\ees
Applying the Laplacian to \eq{BoxPhiPsi} and using \eq{n2PhiPsi} we get 
\be
m_g^2\(\Box-m_g^2\) \Phi=0\, ,
\ee
and therefore,
for $m_g\neq 0$,
we  recover our flat-space result (\ref{BoxPhimm}). 

At first sight \eq{AdPhi} seems to imply that, in de~Sitter,  the radiative degree of freedom in the scalar sector is described by $a^{-2}(\Phi+\Psi)$. The situation is however more subtle. \Eqs{dpsi}{dphi} have been obtained by treating $\{\phi,\psi,\gamma,\lambda\}$ as independent fields. At this level $\Phi$ and $\Psi$ are therefore simply a {\em notation} for the combinations given by \eqs{PhiDS}{PsiDS}. In particular, since $\Psi$ includes a terms $\ddot{\lambda}$, \eq{dlambda} is a fully dynamical equation for the variable $\lambda$, and not a constraint that allows us to eliminate $\lambda$ from the theory
in favor of $\Psi$ and $\Phi$. A way to understand this point is to observe that, in order to have a well-defined Cauchy problem, we must assign the values of $h_{ij}$ and 
$\dot{h}_{ij}$ at some initial time $t_0$. In the scalar sector, this means that we assign $\{\phi(t_0), \lambda(t_0)\}$ and
$\{\dot{\phi}(t_0), \dot{\lambda}(t_0)\}$.
However, for $H\neq 0$, $\Phi$ contains a term $\dot{\lambda}$ and $\Psi$ contains terms $\dot{\lambda}$ and $\ddot{\lambda}$. Therefore, assigning $\lambda(t_0)$ and $\dot{\lambda}(t_0)$ is not sufficient to provide the initial values  
$(\Phi+\Psi)(t_0)$ nor $(\dot{\Phi}+\dot{\Psi})(t_0)$.
 Thus, \eqs{n2PhiPsi}{BoxPhiPsi} are not a closed set of equations that can be solved for $\Phi$ and $\Psi$, once given the initial conditions on the metric (in which case one could have then determined $\lambda$ from \eq{dlambda}). We still need \eq{dlambda}, written as a second-order differential equation in $\lambda$, to evolve the system. 

Thus, these manipulations of the equations of motion are not the correct way of eliminating the non-dynamical variables, and
$(\Phi+\Psi)$ is not the radiative degree of freedom in the scalar  sector. In order to correctly identify the field that describes  the radiative degree of freedom in the scalar  sector we must go through the procedure presented in sect.~\ref{sect:DS}.

\bibliography{myrefs_massive}
\end{document}